\documentclass[aps,prd,eqsecnum,showpacs,amsmath,amssymb,superscriptaddress,nofootinbib]{revtex4-2}

\usepackage[utf8]{inputenc}
\usepackage[english]{babel}
\usepackage{amsmath,amssymb}
\usepackage{amsfonts}
\usepackage{graphicx}
\usepackage{hyperref}
\usepackage{xcolor}
\usepackage{caption}
\usepackage{ulem}
\definecolor{linkcolor}{named}{black}
\hypersetup{ linkcolor=linkcolor, colorlinks=true}
 
\newcommand{\beq}{\begin{equation}}
\newcommand{\eeq}{\end{equation}}
\newcommand{\bea}{\begin{eqnarray}}
\newcommand{\eea}{\end{eqnarray}}
\newcommand{\const}{\text{const}}
\newcommand{\eps}{\varepsilon}

\begin{document}
\title{Image of a wormhole with an arbitrary throat profile}

\author{Valeria A. Ishkaeva}
\email{ishkaeva.valeria@mail.ru}
\affiliation{Institute of Physics, Kazan Federal University, Kremliovskaya str. 16a, Kazan 420008, Russia}

\author{Sergey V. Sushkov}
\email{sergey$_\,$sushkov@mail.ru}
\affiliation{Institute of Physics, Kazan Federal University, Kremliovskaya str. 16a, Kazan 420008, Russia}

\date{\today}

\begin{abstract}
We investigate the observable signatures -- the shadow, the throat silhouette, and the image of a thin accretion disk -- for a family of static, spherically symmetric wormholes with an arbitrary throat profile.
First, we derive expressions for the shadow radius, the throat silhouette radius, and the photon energy shift for a general static, spherically symmetric metric. Then we apply these results to a specific wormhole metric containing three free parameters: the throat radius~$a$, the throat length~$\lambda$, and the parameter~$u_0$ that controls the depth of the gravitational well. We numerically obtain the shadow and silhouette radii as functions of $\lambda$, $u_0$, and $a$, construct accretion disk images for three representative parameter sets, and compare the results with those for a Schwarzschild black hole.
We find that there exist sets of parameters $a$, $\lambda$, $u_0$ such that the wormhole shadow and throat silhouette radii coincide with the shadow and event horizon silhouette of a Schwarzschild black hole of the same mass. Nevertheless, the accretion disk images of these objects differ substantially. In wormhole images, the Doppler effect plays a major role, not the gravitational redshift. As a result, the accreting wormhole images appear brighter.
\end{abstract}

\maketitle

\section{Introduction}

Wormholes are hypothetical spacetime tunnels connecting two distant regions of the same universe, or different universes. As mathematical objects, they were discovered already Wormholes are hypothetical spacetime tunnels connecting two distant regions of the same universe, or different universes. As mathematical objects, they were discovered already in the early days of general relativity, in the fundamental works of Flamm \cite{Flamm}, Einstein and Rosen \cite{EinsteinRosen}, and later Wheeler \cite{Wheeler_Geons,Wheeler_GDs}, who introduced the term ``wormhole''.
There was little interest in these objects until the late 1980s, when Morris and Thorne \cite{MorrisThorne1} demonstrated that traversable wormholes could, in principle, provide interstellar travels. Since then, an extensive literature on wormhole physics has appeared, involving classical solutions, stability analysis, and observational signatures (see, e.g., the reviews \cite{VisserBook,LoboReview,BambiStojkovic,Bronnikov:2023lza}).

Today, the search for astrophysical wormholes has become more active due to significant progress in observations of supermassive compact objects. The Event Horizon Telescope (EHT) collaboration has obtained the first images of the black hole candidates M87$^*$ and Sgr A$^*$~\cite{EHT}, opening a new era of strong‑field gravity tests. Although these observations are agree with the theory of black holes, they do not rule out alternative compact objects, such as wormholes.

Gravitational lensing is one of the most promising methods for detecting wormholes as astrophysical objects. 
Like black holes, wormholes can act as gravitational lenses, bending light rays~\cite{TsuHarYaj,Perlick,NanZhaZak,Muller,Shaikh:2019jfr, Bronnikov:2018nub,Shaikh:2018oul} and producing the \textit{shadow}~\cite{Ishkaeva:2023xny,Pal:2023,Wielgus_etal:2020,Shaikh:2018kfv,Amir:2018pcu,Guerrero:2021,Gjorgjieski:2025uik} if there is a luminous background, and the throat \textit{silhouette} in images of accreting wormholes~\cite{Ishkaeva:2023xny,Gjorgjieski:2025matter,Tan:2025cte,Novikov:2025elo,Macedo:2025ipc,Bambhaniya:2021ugr}.

The two features have different physical origins and, in general, different sizes. The shadow is a dark area in the observer's sky formed by photons that are captured by the object's gravitational field; its boundary is determined by the circular photon orbits -- the photon sphere~\cite{BardeenPress,PerlickTsupko}. The silhouette of the event horizon (or the throat, in the case of a wormhole) is a different feature: it is formed by photons emitted near the compact object and reaching a distant observer after being strongly lensed. In Ref.~\cite{Dokuchaev:2019jqq}, it was shown that in the first EHT image of M87$^*$ the observed dark spot is the silhouette of the event horizon, not the classical shadow.

For wormholes, the situation is more difficult because they do not possess an event horizon. As a result, wormhole image structure can be more difficult.
For example, it has been demonstrated that asymmetric thin-shell wormholes can exhibit double shadows and extra photon rings~\cite{Wielgus_etal:2020,Guerrero:2021}, while images of some wormholes can closely resemble those of Schwarzschild black holes~\cite{Pal:2023,Bambhaniya:2021ugr}. These results show both the potential and the difficulty of distinguishing wormholes from black holes.

Although there are many works devoted to wormhole lensing and shadows, most of them consider wormholes with relatively short throats. Configurations with long throats  have been studied much less. A wormhole metric containing a long throat was introduced in Ref.~\cite{SushkovDwalls}, where it was shown that such a geometry appear naturally in the presence of a spherical domain wall localized near the throat. But gravitational lensing and shadow characteristics of long-throated wormholes have not been studied. It is also worth noting that long-throat wormholes have been studied within the semiclassical theory of gravity, where the vacuum fluctuations of quantized fields serve as the source of spacetime curvature. 
In Ref.~\cite{Popov:2018} it was shown that self-consistent semiclassical Einstein equations admit solutions describing an elongated wormhole throat supported by the vacuum polarization of a quantized scalar field and an electrostatic field.

In this paper we investigate the observable signatures -- the shadow, the throat silhouette, and the image of a thin accretion disk -- for a family of static, spherically symmetric wormholes with an adjustable throat length. As an example, we considered a wormhole, the metric functions of which were taken from Refs.~\cite{Bronnikov:2018nub,SushkovDwalls}. Our wormhole metric functions have two key parameters: $\lambda$, determining the length of the throat, and $u_0$, controlling the depth of the gravitational well. In addition, we have the throat radius $a$. Our goal is to understand how these parameters affect the observable features and to determine whether a wormhole with a long throat can be distinguished from a Schwarzschild black hole.

The article is organized as follows. In Sec.~II we introduce the general static, spherically symmetric wormhole metric and derive the basic equations for geodesic motion, photon spheres, shadows, throat silhouettes, and the energy shift of photons emitted by an accretion disk. Section~III is devoted to a specific two-parameter family of metric functions, its physical interpretation, and the numerical results for the shadow, throat silhouette, and accretion disk image and compare them with the Schwarzschild black hole case. Our conclusions are summarized in Sec.~IV.

\section{General framework: static spherically symmetric wormholes}


\subsection{Wormhole metric}

We adopt a general parametrization for a static, spherically symmetric wormhole spacetime.  
Using the proper radial coordinate $u$, which ranges from $-\infty$ to $+\infty$, the metric can be written as
\begin{equation}
ds^2 = -N^2(u)\,dt^2 + du^2 + r^2(u)\left(d\theta^2 + \sin^2\theta\, d\phi^2\right).
\label{metric_GS}
\end{equation}
In the asymptotic region containing the observer ($u \to +\infty$) the spacetime is required to be flat, which imposes the boundary conditions
\begin{equation}
\lim_{u\rightarrow+\infty} N^2(u) = 1 - \frac{2m}{u}+ O(u^{-2}), \qquad
\lim_{u\rightarrow+\infty} r^2(u) = u^2(1+ O(u^{-1})), \label{flat_conditions}
\end{equation}
where the constant $m$ is the mass of the wormhole as measured by a distant observer. Also, $N(u)$ must be finite everywhere to avoid event horizons or curvature singularities.

Without loss of generality we locate the wormhole throat at $u_{\text{th}} = 0$, so that positive and negative values of $u$ correspond to the two distinct sides of the throat.  
Therefore, the function $r(u)$ must have the global minimum at $u_{\text{th}}=0$:
\begin{equation}
r'(0) = 0, \qquad r''(0) \ge 0,\label{global_minimum}
\end{equation}
so that $r_{\text{th}}=r(0)$ is the radius of the wormhole throat.

It is worth remarking that the region $u \to -\infty$, which lies beyond the throat and contains no observer, is not required to obey the asymptotically flat conditions stated above.  
A well‑known illustration is provided by the Ellis–Bronnikov wormhole: in that solution the two asymptotic limits are $\lim_{u\to+\infty} N^2(u) = 1 - 2m/u$ on the observer's side, whereas on the opposite side one finds $\lim_{u\to-\infty} N^2(u) = e^{-2\pi m/a}\,(1 + 2m/|u|)$.

\subsection{Embedding diagram}

An embedding diagram provides a standard and intuitive way to visualize the spatial geometry of a wormhole and the geodesic motion of test particles within it.  
To construct such a diagram, we consider a two-dimensional spatial section of the spacetime (\ref{metric_GS}) obtained by fixing the time coordinate $t = \const$ and restricting to the equatorial plane $\theta = \pi/2$:
\begin{equation}
ds^2_{(2)} = du^2 + r^2(u)\, d\phi^2.
\label{eq:2D_slice}
\end{equation}
We then suppose that this surface is embedded into a three-dimensional Euclidean space with the metric given in cylindrical coordinates:
\begin{equation}
ds^2_{\text{Eucl}} = dz^2 + dR^2 + R^2\, d\phi^2.
\end{equation}
An axially symmetric embedded surface can be described by a function $z = z(R)$, so that the metric induced on this surface becomes
\begin{equation}
ds^2_{\text{ind}} = \bigl(z'^2_R + 1\bigr)\, dR^2 + R^2\, d\phi^2,
\label{eq:induced_metric}
\end{equation}
where $z'_R \equiv dz/dR$.
Comparing the two metrics (\ref{eq:2D_slice}) and (\ref{eq:induced_metric}), we obtain
\begin{equation}
R(u) = r(u), \qquad \frac{dz}{du} = \pm \sqrt{1 - r'^2(u)}.
\label{eq:diagram}
\end{equation}
Examples of embedding diagrams for the specific wormhole model considered in this work are presented in Fig.~\ref{diags}.

\subsection{Geodesic motion}

The trajectories of test particles and light rays are governed by the Hamilton–Jacobi equation
\begin{equation}
	g^{i j}\frac{\partial S}{\partial x^{i}}\frac{\partial S}{\partial x^{j}}=-\mu^2,\label{HY_GS}
\end{equation}
where $\mu$ is the rest mass of the particle ($\mu=0$ for photons). Because the metric (\ref{metric_GS}) admits two Killing vectors, $\partial_t$ and $\partial_\phi$, the action $S$ can be separated as \cite{Carter}
\begin{equation}
	S=-Et+L\varphi+S_r(r)+S_\theta(\theta),\label{S_GS}
\end{equation}
where $E\equiv-p_t$ is the total energy and $L\equiv~p_{\phi}$ is the azimuthal angular momentum. 
Thus, from equations (\ref{HY_GS}) and (\ref{S_GS}) we obtain
\begin{equation}
	-r^2(u)\left(\frac{dS_r}{dr}\right)^2+\frac{E^2r^2(u)}{N^2(u)}-\mu^2r^2(u)=\left(\frac{dS_\theta}{d\theta}\right)^2+\frac{L^2}{\sin^2\theta}
	\equiv K,\nonumber
\end{equation}
where $K$ is the square of the total angular momentum. 
Using the equality
\begin{equation}
	\frac{dx^{i}}{d\lambda}=g^{ij}\frac{\partial S}{\partial x^{j}},\nonumber
\end{equation}
where $\lambda$ is an affine parameter, we can write the equations of motion in the following form
\begin{eqnarray}
	\frac{dt}{d\lambda}&=&\frac{E}{N^2(u)},\label{t_GS}\\
	\frac{du}{d\lambda}&=&\pm E\sqrt{\frac{1}{N^2(u)}-\frac{k^2}{r^2(u)}-\frac{1}{\eps^2}}=\pm E\sqrt{U(u)},\label{u_GS}\\
    \frac{d\theta}{d\lambda}&=&\pm\frac{E}{r^2(u)}\sqrt{k^2-\frac{l^2}{\sin^2\theta}}=\pm\frac{E}{r^2(u)}\sqrt{\Theta(\theta)},\label{theta_GS}\\
	\frac{d\phi}{d\lambda}&=&E\frac{l}{r^2(u)\sin^2\theta}.\label{phi_GS}
\end{eqnarray} 
Here we have introduced the effective potentials $U(u)$ and $\Theta(\theta)$, as well as the dimensionless impact parameters
\begin{equation}
    \eps = \frac{E}{\mu}, \qquad l = \frac{L}{E}, \qquad k = \frac{\sqrt{K}}{E}.
\end{equation}
For massless particles ($\mu=0$) the term $1/\eps^2$ in (\ref{u_GS}) vanishes; consequently, photon trajectories are determined by only two impact parameters, $l$ and $k$.

\subsubsection{Photon trajectories}
\label{sec:circ_photons}

We now focus on photon trajectories, supposing $\mu = 0$.  
Analyzing the radial equation (\ref{u_GS}), we conclude that there are three distinct types of photon trajectories:
\begin{itemize}
    \item[(i)] Trajectories that possess a turning point where $U(u) = 0$; such rays approach the wormhole, are deflected around it, and escape back to infinity.
    \item[(ii)] Trajectories with no turning point, for which $U(u) > 0$ everywhere; these rays cross the throat and emerge into the opposite region of spacetime.
    \item[(iii)] Trajectories that infinitely twisting around the wormhole, satisfying $U(u) = 0$ and $U'(u) = 0$ \cite{BardeenPress}; these correspond to circular photon orbits.
\end{itemize}
The circular photon orbits are the boundary between the other two types of photon trajectories.
Consequently, if a luminous background is placed behind the wormhole, photons of type (i) form an image of that background, photons of type (ii) constitute the wormhole shadow, and the critical rays of type (iii) define the boundary of the shadow.

Since the space-time of the wormhole (\ref{metric_GS}) is spherically symmetric, without loss of generality we will assume that photons move in the equatorial plane $\theta = \pi/2$. 
In this plane $\Theta(\theta)=\const$, and the impact parameters become related by $k^2 = l^2$.
The position $u_{\text{ph}}$ of a circular photon orbit and the corresponding impact parameter $l_{\text{ph}}$ are determined by the simultaneous conditions \cite{BardeenPress}
\begin{equation}
    U(u_{\text{ph}}) = 0, \qquad U'(u_{\text{ph}}) = 0.
\end{equation}
Substituting the explicit form of $U(u)$ from (\ref{u_GS}), we obtain
\begin{eqnarray}
    U(u_{\text{ph}}) = 0: &\quad& \left.\left(\frac{1}{N^2} - \frac{l_{\text{ph}}^2}{r^2}\right)\right|_{u=u_{\text{ph}}} = 0 \quad \Rightarrow \quad l_{\text{ph}}^2 = \left.\frac{r^2}{N^2}\right|_{u=u_{\text{ph}}}, \label{Uph_GS} \\
    U'(u_{\text{ph}}) = 0: &\quad& \left.\left(\frac{2 l^2_{\text{ph}}\, r'}{r^3} - \frac{2 N'}{N^3}\right)\right|_{u=u_{\text{ph}}} = 0 \quad \Rightarrow \quad \left.\left(\frac{r'}{r} - \frac{N'}{N}\right)\right|_{u=u_{\text{ph}}} = 0. \label{U'ph_GS}
\end{eqnarray}
The circular orbit is unstable if $U''(u_{\text{ph}}) > 0$.  
Evaluating the second derivative and using the conditions (\ref{Uph_GS})--(\ref{U'ph_GS}) to eliminate $l^2$, we find
\begin{equation}
    U''(u) = \frac{r''}{r} - \frac{3r'^2}{r^2} - \frac{N''}{N} + \frac{3N'^2}{N^2}
    \quad \longrightarrow \quad
    U''(u_{\text{ph}}) = \frac{r''(u_{\text{ph}})}{r(u_{\text{ph}})} - \frac{N''(u_{\text{ph}})}{N(u_{\text{ph}})}.
    \label{U''ph_GS}
\end{equation}

Further detailed discussion of photon orbits and strong gravitational lensing by wormholes can be found in Refs.~\cite{Shaikh:2018oul,Shaikh:2019jfr}, where the authors show that the throat itself may act as an effective photon sphere, and that photon and antiphoton spheres can appear on both sides of the throat, leading to novel gravitational lensing signatures distinct from those of black holes.

\subsubsection{Massive particle trajectories}

Now, we consider the motion of massive test particles, for which $\mu \neq 0$.
Of primary interest for our purposes is the existence of stable circular orbits, since they provide the foundation for a thin accretion disk.  
For this reason we restrict our analysis to the equatorial plane of the wormhole, supposing $\theta = \pi/2$ and consequently $k^2 = l^2$.

A circular orbit of a massive particle is determined by the condition that the radial effective potential $U(u)$ and its first derivative vanish simultaneously~\cite{BardeenPress}:
\begin{equation}
    U(u) = 0, \qquad U'(u) = 0.
\end{equation}
Stability of the orbit further requires that the second derivative be non‑positive:
\begin{equation}
    U''(u) \le 0.
\end{equation}
Substituting the explicit form of $U(u)$ from Eq.~(\ref{u_GS}) into these conditions, we obtain
\begin{eqnarray}
    U = 0: &\quad& \frac{1}{N^2} - \frac{l^2}{r^2} - \frac{1}{\eps^2} = 0, \\
    U' = 0: &\quad& \frac{l^2 r'}{r^3} - \frac{N'}{N^3} = 0, \label{U'part_GS} \\
    U'' \le 0: &\quad& \frac{l^2}{r^3}\left(r'' - \frac{3r'^2}{r}\right) - \frac{1}{N^3}\left(N'' - \frac{3N'^2}{N}\right) \le 0. \label{U''part_GS}
\end{eqnarray}

Assuming $N(u) \neq \const$, the first two equations can be solved for the impact parameters associated with the circular orbit $u$:
\begin{equation}
    l_{\text{orbit}}^2 = \frac{r^3 N'}{N^3 r'}, \qquad
    \eps_{\text{orbit}}^2 = \frac{r^2 N^2}{r^2 - l_{\text{orbit}}^2 N^2} = \frac{N^3 r'}{N r' - r N'}.
    \label{lpart_GS}
\end{equation}
The boundary between stable and unstable circular motion defines the innermost stable circular orbit (ISCO).  
At the ISCO, the second derivative vanishes, $U''(u_{\text{isco}}) = 0$, which leads to the equation
\begin{equation}
    \left.\left[ N'\left(\frac{r''}{r'} - \frac{3r'}{r}\right) + \frac{3N'^2}{N} - N'' \right]\right|_{u = u_{\text{isco}}} = 0.
    \label{uisco_GS}
\end{equation}

\subsubsection{Orbits exactly at the throat}

Circular orbits located exactly at the wormhole throat $u_{\text{th}} = 0$ represent a special case that deserves separate consideration.  
By definition, the throat is a minimum of $r(u)$, so that
\begin{equation}
    r'(0) = 0, \qquad r''(0) \ge 0.
\end{equation}
Substituting $r'(0) = 0$ into the general condition for circular orbits, Eq.~(\ref{U'ph_GS}) for photons and Eq.~(\ref{U'part_GS}) for massive particles, we immediately find that the existence of such orbits requires
\begin{equation}
    N'(0) = 0.
\end{equation}
In other words, a circular orbit can reside on the throat only if the function $N(u)$ is symmetric with respect to $u = 0$ (at least locally).

When this symmetry condition is fulfilled, the impact parameters for particles moving on the throat orbit are given by
\begin{eqnarray}
    \text{photons:} &\quad& l_{\text{th}} = \pm \frac{r_{\text{th}}}{N(0)}, \\
    \text{massive particles:} &\quad& \frac{l_{\text{th}}^2}{r_{\text{th}}^2} + \frac{1}{\eps_{\text{th}}^2} = \frac{1}{N^2(0)}.
\end{eqnarray}

The stability of the throat orbit is determined by the sign of the second derivative $U''(0)$.  
For photons, evaluating the general expression (\ref{U''ph_GS}) at $u = 0$ and using $N'(0) = 0$ yields
\begin{equation}
    U''_{\text{ph}}(0) = \frac{r''(0)}{r_{\text{th}}} - \frac{N''(0)}{N(0)}.
    \label{U''_th_ph}
\end{equation}
For massive particles, the corresponding expression follows from Eq.~(\ref{U''part_GS}) together with the throat conditions:
\begin{equation}
    U''_{\text{massive}}(0) = \frac{l_{\text{th}}^2\,r''(0)}{r^3(0)}-\frac{N''(0)}{N^3(0)}.
    \label{U''_th_mass}
\end{equation}
If $U''(0) < 0$, the orbit is stable; if $U''(0) > 0$, it is unstable.  
Note that for massive particles the stability depends not only on the local curvature of $r(u)$ and $N(u)$, but also on the particle's energy $\eps_{\text{th}}$, whereas for photons it is purely a geometric property.

A detailed discussion of circular orbits and photon spheres at wormhole throats can be found in Ref.~\cite{Gjorgjieski:2025uik}.

\subsection{Wormhole shadow}\label{subsec:shadow}

To determine the apparent size of the wormhole shadow as seen by a distant observer, we must project the critical photon trajectories onto the observer's sky.  
The celestial coordinates $(\alpha, \beta)$ of an incoming light ray are given by the standard relations~\cite{Vasquez}
\begin{equation}
    \alpha = -r_o^2 \sin\theta_o \left.\frac{d\phi}{dr}\right|_{r_o}, \qquad
    \beta = r_o^2 \left.\frac{d\theta}{dr}\right|_{r_o},
    \label{skycoor}
\end{equation}
where $r_o$ and $\theta_o$ denote the observer's spatial coordinates.

As discussed in Sec.~\ref{sec:circ_photons}, the boundary of the wormhole shadow is formed by photons moving in the circular orbit.
Since the spacetime is spherically symmetric, we may place the observer in the equatorial plane, $\theta_o = \pi/2$, without loss of generality. In this case the shadow is a circle with the radius $\alpha_{\text{sh}}$.
Taking the limit $r_o \to \infty$ (distant observer) and using the equations of motion (\ref{t_GS})--(\ref{phi_GS}), we obtain
\begin{equation}
    \alpha_{\text{sh}} = |l_{\text{ph}}| = \left|\frac{r(u_{\text{ph}})}{N(u_{\text{ph}})}\right|.
    \label{shadow_radius}
\end{equation}

It is worth noting that some wormhole spacetimes can possess more than one photon sphere. 
For instance, the throat itself can act as an effective photon sphere in addition to the usual one present outside the throat~\cite{Shaikh:2018oul,Shaikh:2019jfr}. 
In such cases, the shadow boundary is determined by the outermost photon sphere located on the observer's side, as the shadow is formed by rays originating from a luminous background situated in the same asymptotic region as the observer. 
Inner photon spheres, if present, do not contribute to the shadow boundary but can produce additional relativistic Einstein rings or bright features inside the shadow region~\cite{Shaikh:2018oul,Wielgus_etal:2020,Guerrero:2021}. 
These internal structures may serve as distinctive observational signatures that help differentiate wormholes from black holes.

\subsection{Silhouette of the throat}\label{subsec:silhouette}

In addition to the wormhole shadow, whose boundary is formed by photons moving in a circular orbit, another distinct dark feature that we can observe is the silhouette of the throat.  
The boundary of this silhouette is formed by photons that are emitted directly from the wormhole throat. It is typically visible in simulated images of accretion disks surrounding compact objects, where it appears as a dark central spot distinct from the photon sphere shadow.
In the context of black hole imaging, such a feature is naturally expected, as no light can escape from the event horizon.  
For wormholes, however, the throat is in principle traversable, and its silhouette is not a priori obvious.  
Nevertheless, knowing the size of the throat silhouette is valuable, especially for asymmetric configurations such as the Ellis–Bronnikov wormhole~\cite{Ishkaeva:2023xny}.  

Following the approach of Ref.~\cite{Dokuchaev:2019jqq}, we now derive an integral equation that determines the radius of this silhouette as seen by a distant observer.

The trajectories of photons that form the direct image of a source (i.e., rays that do not cross the equatorial plane between the source and the observer) are governed by the integral relations
\begin{equation}
\int_{u_s}^{u_o} \frac{du}{r^2(u) \sqrt{N^2(u) - k^2/r^2(u)}} =
\int_{\theta_o}^{\theta_s} \frac{d\theta}{\sqrt{k^2 - l^2/\sin^2\theta}}.
\label{noturnpoint}
\end{equation}
If the ray possesses a turning point in the polar direction, the right‑hand side splits into two integrals:
\begin{equation}
\int_{u_s}^{u_o} \frac{du}{r^2(u) \sqrt{N^2(u) - k^2/r^2(u)}} =
\int_{\theta_{\min}}^{\theta_s} \frac{d\theta}{\sqrt{k^2 - l^2/\sin^2\theta}} +
\int_{\theta_{\min}}^{\theta_o} \frac{d\theta}{\sqrt{k^2 - l^2/\sin^2\theta}},
\label{turnpoint}
\end{equation}
where $\theta_{\min} = \arcsin(|l|/k)$ is the turning point determined by $\Theta(\theta) = 0$.

For the silhouette of the throat, we consider a photon emitted from the throat itself ($u_s = 0$) and reaching a distant observer located in the equatorial plane ($u_o \to \infty$, $\theta_o = \pi/2$).  
The boundary of the silhouette corresponds to the critical ray that has a turning point in the polar $\theta$-direction. Setting $\theta_s = \pi/2$ and using the symmetry of the configuration, Eq.~(\ref{turnpoint}) reduces to
\begin{equation}
\int_0^\infty \frac{du}{r^2(u) \sqrt{N^2(u) - k^2/r^2(u)}} =
2 \int_{\theta_{\min}}^{\pi/2} \frac{d\theta}{\sqrt{k^2 - l^2/\sin^2\theta}}.
\label{silint}
\end{equation}
From the celestial coordinate expressions (\ref{skycoor}), taking the limit $r_o \to \infty$ and $\theta_o = \pi/2$, the observed radius of the throat silhouette is $\alpha_{\text{sil}} = k$.  Thus, taking the integral on the right side of (\ref{silint}), we obtain the final equation for the radius of the throat silhouette:
\begin{equation}
\int_0^\infty \frac{du}{r^2(u) \sqrt{N^2(u) - \alpha_{\text{sil}}^2/r^2(u)}} = \frac{\pi}{\alpha_{\text{sil}}}.
\label{sileq}
\end{equation}
Equation (\ref{sileq}) must be solved numerically for a given wormhole metric.

It is worth noting that, in contrast to the shadow radius $\alpha_{\text{sh}}$, which is determined only by the location of the photon sphere, the silhouette radius $\alpha_{\text{sil}}$ depends on the shape of the functions $N(u)$ and $r(u)$.

\subsection{Image of an accretion disk}\label{sec:accretion}

In this section we outline a simplified procedure for constructing the image of a thin accretion disk surrounding the wormhole, closely following the approach of Ref.~\cite{Dokuchaev:2019jqq}.  
We restrict our attention to photons emitted by the infalling matter that resides inside the innermost stable circular orbit, i.e. in the region $u \in [0, u_{\text{isco}}]$.  
Only two physical effects are taken into account when computing the observed flux: the gravitational redshift and the Doppler shift.  
To evaluate these shifts conveniently, we employ the locally non‑rotating reference frame (LNRF).

In any stationary spherically symmetric asymptotically flat spacetime, one can introduce locally nonrotating reference frames (LNRFs) \cite{BardeenPress} in which observers are moving along the world lines $u = const, \theta = const, \varphi = const$.
The orthonormal tetrad carried by such an observer for the metric (\ref{metric_GS}) is given by
\begin{equation}
\mathbf{e}_{(t)} = \frac{1}{N}\partial_t,\quad
\mathbf{e}_{(u)} = \partial_u,\quad
\mathbf{e}_{(\theta)} = \frac{1}{r}\partial_\theta,\quad
\mathbf{e}_{(\phi)} = \frac{1}{r\sin\theta}\partial_\phi.\label{LNRF_basis}
\end{equation}


Consider a small element of the accretion flow moving with azimuthal velocity $v^{(\phi)}$ and radial velocity $v^{(u)}$ relative to the LNRF.  
Using the four‑velocity components derived from Eqs.~(\ref{t_GS})--(\ref{phi_GS}), the LNRF velocities are
\begin{eqnarray}
    v^{(\phi)} &=& \frac{L\,N(u)}{E\,r(u)}, \\[2mm]
    v^{(u)}   &=& \frac{N(u)}{E}\sqrt{\frac{E^2}{N^2(u)} - \frac{L^2}{r^2(u)} - \mu^2}.
\end{eqnarray}
For a photon with impact parameters $l$ and $k$, the LNRF components of its four‑momentum are obtained by projecting the coordinate components onto the tetrad (\ref{LNRF_basis}):
\begin{eqnarray}
    p^{(t)} &=& \frac{1}{N(u)}, \\[2mm]
    p^{(u)} &=& \sqrt{\frac{1}{N^2(u)} - \frac{k^2}{r^2(u)}}, \\[2mm]
    p^{(\phi)} &=& \frac{l}{r(u)}.
\end{eqnarray}
Then the photon energy in the comoving frame of this fragment is \cite{Dokuchaev:2019jqq}
\begin{equation}
    \epsilon(l,k) = \frac{p^{(t)} - v^{(\phi)} p^{(\phi)} - v^{(u)} p^{(u)}}{\sqrt{1 - \bigl(v^{(\phi)}\bigr)^2 - \bigl(v^{(u)}\bigr)^2}}.
    \label{photon_energy}
\end{equation}
The corresponding frequency shift factor (the ratio of the detected frequency to the emitted frequency) is $g(l,k) = 1/\epsilon(l,k)$.


The construction of a simplified accretion disk image proceeds as follows:
\begin{enumerate}
    \item For a given emission radius $u_s \in [0, u_{\text{isco}}]$ (with $\theta_s = \pi/2$), we solve the integral equations of motion (\ref{noturnpoint}, \ref{turnpoint}) to find the impact parameters $l$ and $k$ of photons that reach a distant observer located at $(u_o \to \infty, \theta_o)$.
    \item Using the obtained parameters $l$ and $k$, the celestial coordinates $(\alpha, \beta)$ of the photon on the observer's sky are computed from Eq.~(\ref{skycoor}).
    \item For each photon trajectory, the energy shift factor $g(l,k)$ is obtained from the Eq.~(\ref{photon_energy}).
    \item The final image is constructed by assigning the corresponding intensity to each point $(\alpha, \beta)$ on the observer's sky.
\end{enumerate}


\section{A massive wormhole with a long throat}

\subsection{Preliminary considerations: massless wormholes}

Before specifying a concrete model, it is instructive to examine the simplest possible wormhole configuration with $N(u) = 1$.  In this case, the effective radial potential for massive particles reduces to
\begin{equation}
    U(u) = 1 - \frac{l^2}{r^2(u)} - \frac{1}{\eps^2}.
\end{equation}
The conditions for circular orbits, $U(u)=0$ and $U'(u)=0$, immediately require $r'(u)=0$.  
Hence any circular orbit must be located at the throat $u=0$.  
Evaluating the second derivative at the throat gives
\begin{equation}
    U''(0) = \frac{2 l^2 r''(0)}{r^3(0)} > 0,
\end{equation}
which shows that such orbits are always unstable.  
Consequently, a massless wormhole cannot support a thin accretion disk because there are no stable circular orbits of massive particles. Therefore, we will consider \textit{massive} wormholes with a non‑trivial function $N(u)$.


\subsection{Choice of metric functions}

We now consider a specific wormhole model by choosing the metric functions as follows:
\begin{eqnarray}
    N(u) &=& \exp\left[m\left(\arctan\sqrt{u^2+u_0^2} - \frac{\pi}{2}\right)\right], \label{N_choice} \\
    r(u) &=& u \coth\left(\frac{u}{\lambda}\right) - \lambda + a. \label{r_choice}
\end{eqnarray}
The function $N(u)$ is based on the form proposed in Ref.~\cite{Bronnikov:2018nub}, but we have introduced an additional parameter $u_0$ to make the function symmetric with respect to the throat at $u=0$.  
For large $|u|$, this function behaves as
\begin{equation}
    N^2(u) \simeq 1 - \frac{2m}{|u|} + \mathcal{O}\!\left(|u|^{-2}\right),
\end{equation}
from which it is clear that $m$ is the gravitational mass of the wormhole measured by a distant observer.

The function $r(u)$ is taken from Ref.~\cite{SushkovDwalls}, where it was shown that $a$ is the radius of the wormhole throat and the parameter $\lambda$ characterizes the length of the throat.  
This function satisfies the asymptotic flatness conditions (\ref{flat_conditions}) and possesses a global minimum at $u=0$, as required by (\ref{global_minimum}).

\subsection{Embedding diagrams}

\begin{figure}[h!]
	\begin{minipage}[h]{0.49\linewidth}
		\center{\includegraphics[width=1\linewidth]{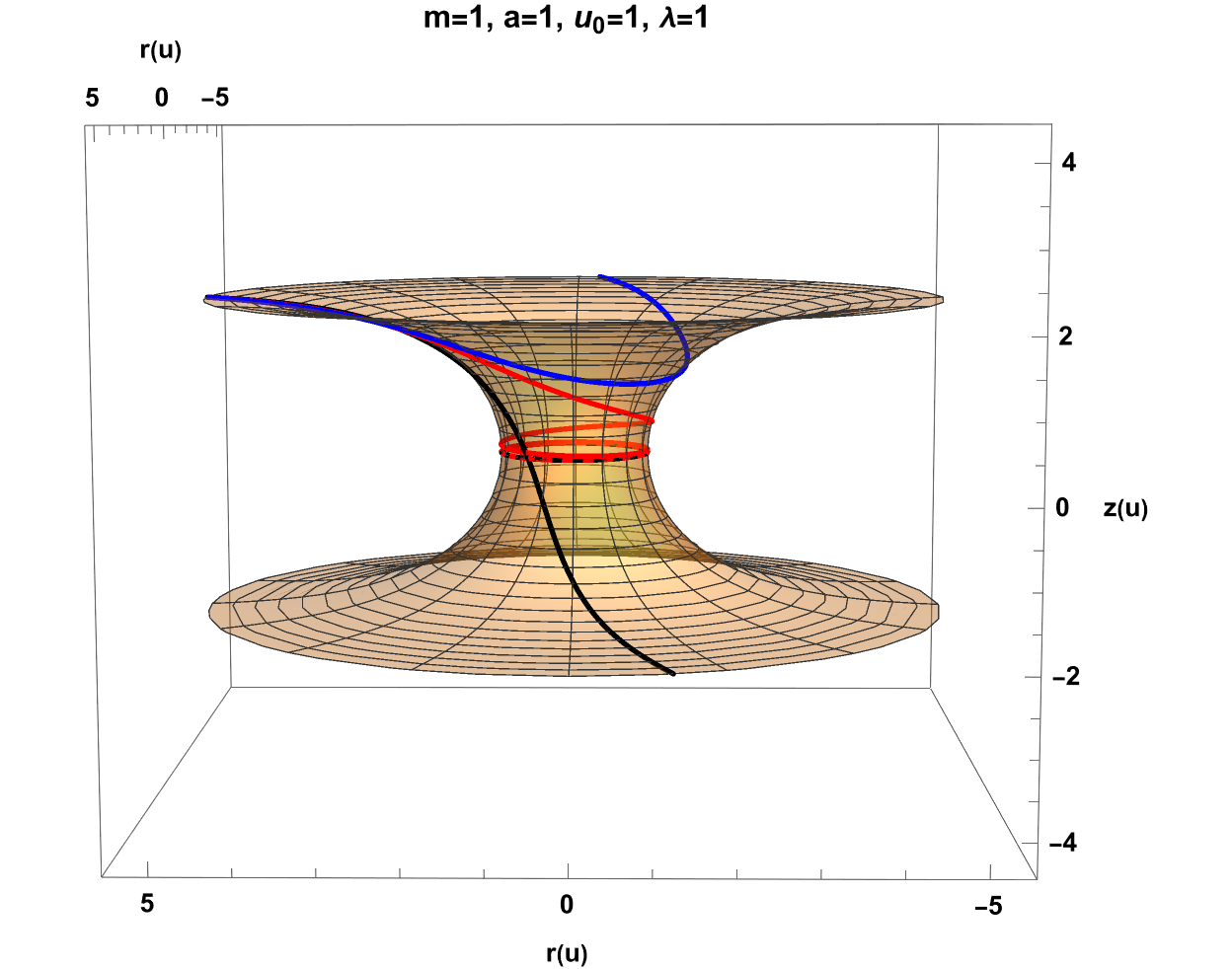}}
	\end{minipage}
	\hfill
	\begin{minipage}[h]{0.49\linewidth}
		\center{\includegraphics[width=1\linewidth]{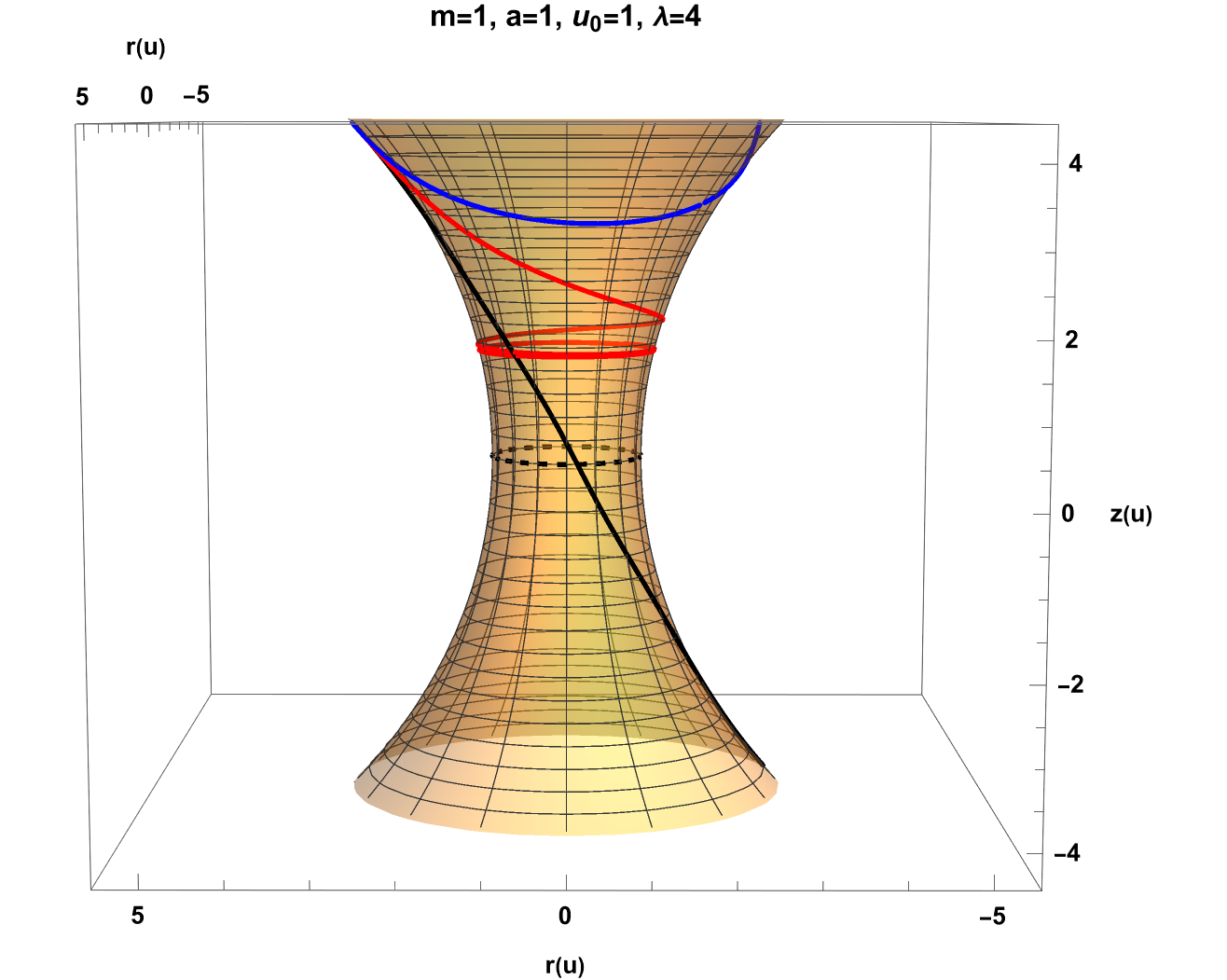}}
	\end{minipage}
	\vfill
	\begin{minipage}[h]{0.49\linewidth}
		\center{\includegraphics[width=1\linewidth]{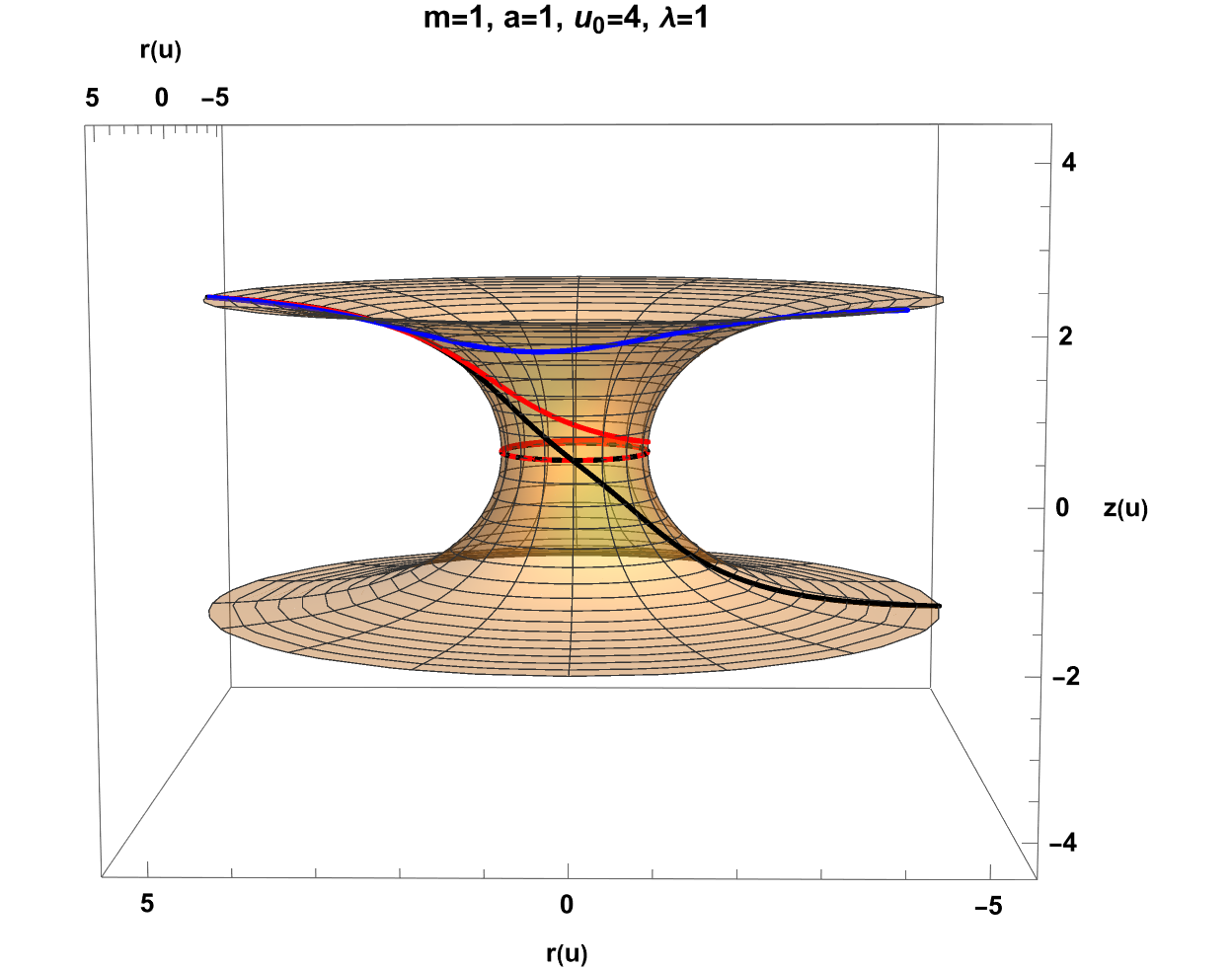}}
	\end{minipage}
	\hfill
	\begin{minipage}[h]{0.49\linewidth}
		\center{\includegraphics[width=1\linewidth]{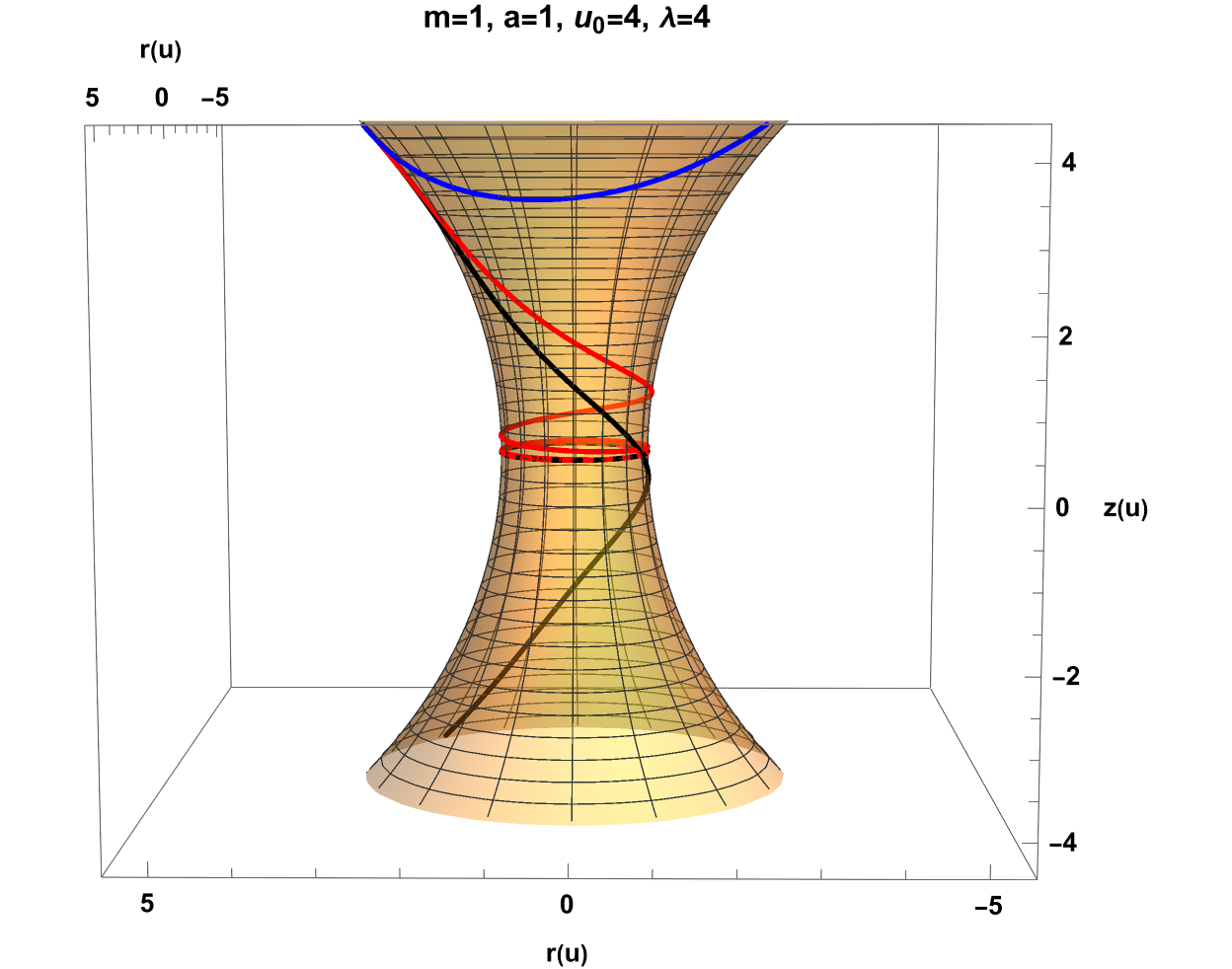}}
	\end{minipage}
	\caption{\label{diags}Embedding diagrams of the wormhole for different values of the throat length parameter $\lambda$ and the parameter $u_0$. Three types of photon trajectories are also shown: black trajectories ($|l|>|l_{\text{ph}}|$) cross the throat, blue trajectories ($|l|<|l_{\text{ph}}|$) bend around the wormhole, and red trajectories correspond to photons that move in circular orbits; the black dotted circle is the throat. ($|l|=|l_{\text{ph}}|$).}
\end{figure}

Embedding diagrams provide a convenient visualisation of the spatial geometry of the wormhole for different values of the parameters $\lambda$ and $u_0$.  
Figure~\ref{diags} shows several examples of such diagrams, constructed using Eq.~(\ref{eq:diagram}) for the chosen metric functions (\ref{N_choice})--(\ref{r_choice}).

On each diagram we have superimposed several photon trajectories.  
The black curves correspond to photons with an impact parameter $l=1 > l_{\text{ph}}$; these rays have no turning point, cross the throat, and emerge on the other side of the wormhole.  
The blue curves represent photons with $l=2.5 <l_{\text{ph}}$, which possess a turning point and are deflected back to the same asymptotic region from which they came.  
The red curves show the critical rays with $l_{\text{ph}}\approx 2.2$, which asymptotically approach the circular photon orbit.

By comparing the panels in Fig.~\ref{diags}, one can clearly see how the parameter $\lambda$ affects the shape of the throat.  
For small $\lambda$ the throat is sharply curved, whereas for large $\lambda$ the central region becomes nearly cylindrical, reflecting the long throat geometry.  

The parameter $u_0$ does not influence the spatial geometry of the wormhole, but it affects photon trajectories, as will be shown in the next section.

\clearpage
\subsection{Photon trajectories and the circular orbit}

\begin{figure}[h!]
    \begin{minipage}[h]{0.48\linewidth}
        \center{\includegraphics[width=1\linewidth]{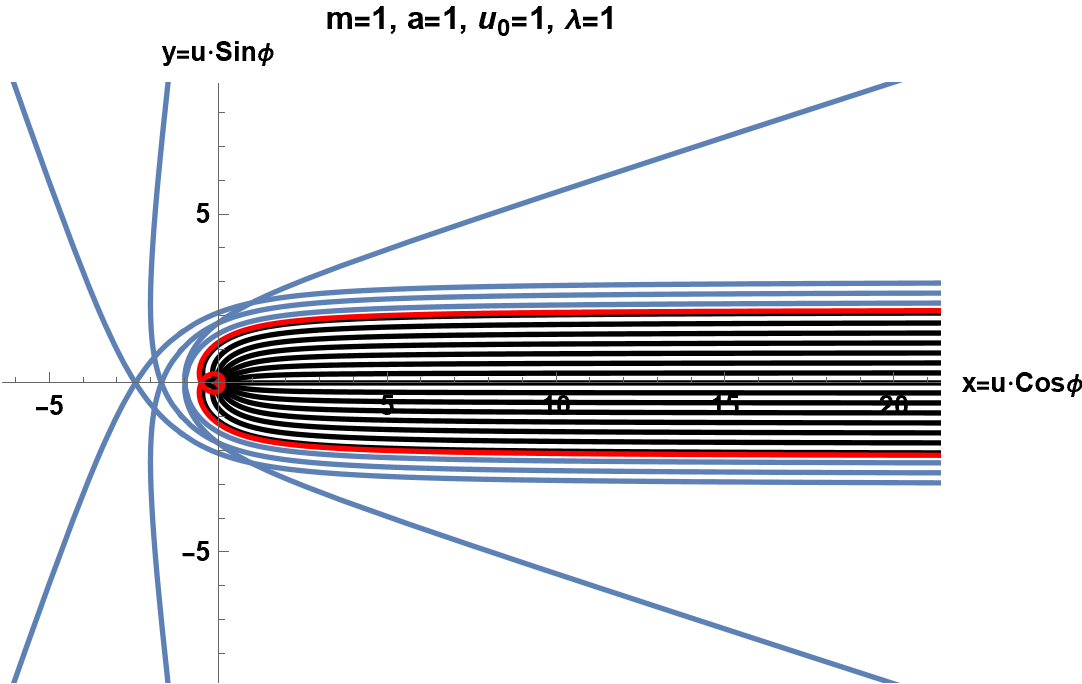}}
    \end{minipage}
    \hfill
    \begin{minipage}[h]{0.48\linewidth}
        \center{\includegraphics[width=1\linewidth]{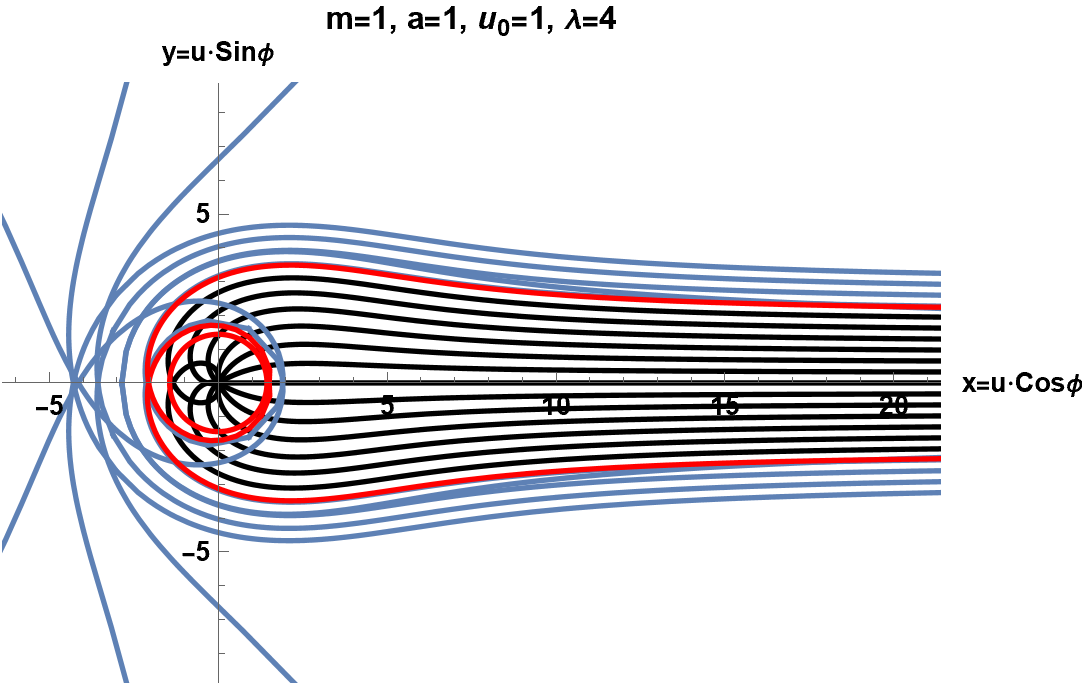}}
    \end{minipage}
    \vfill
    \begin{minipage}[h]{0.48\linewidth}
        \center{\includegraphics[width=1\linewidth]{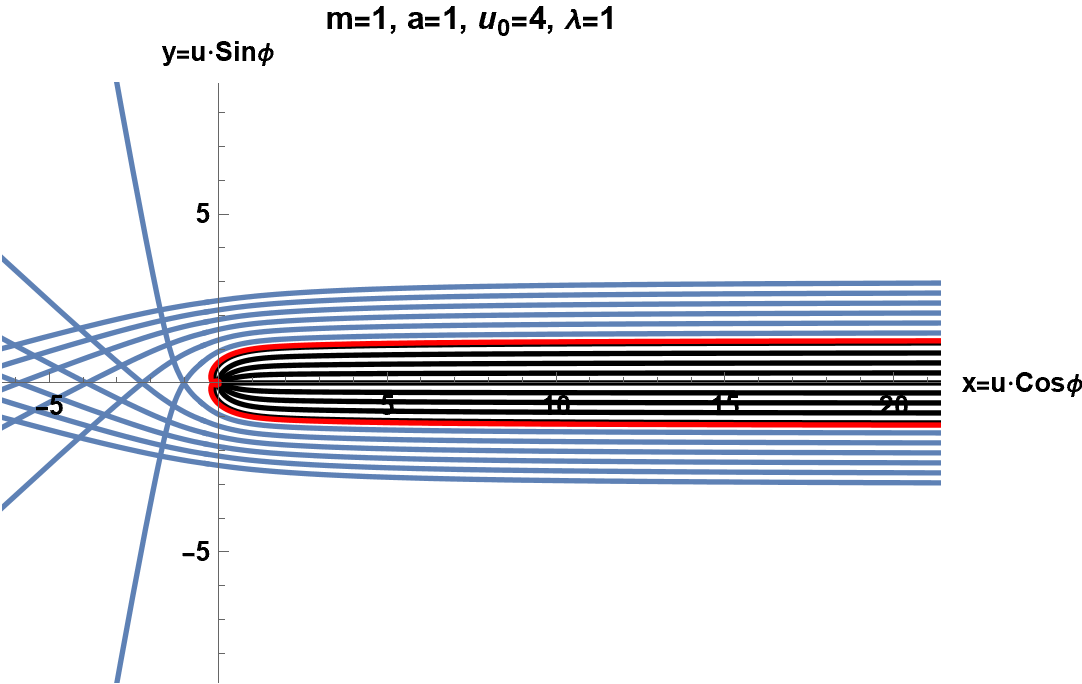}}
    \end{minipage}
    \hfill
    \begin{minipage}[h]{0.48\linewidth}
        \center{\includegraphics[width=1\linewidth]{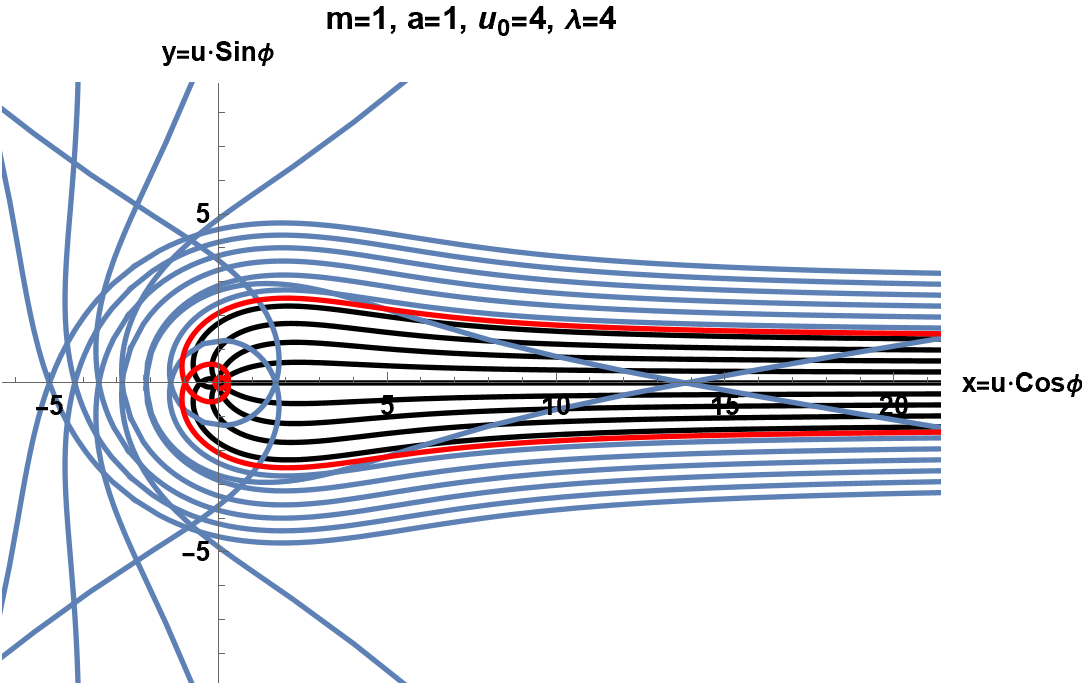}}
    \end{minipage}
    \caption{\label{AllTraj}Photon trajectories $u(\phi)$ in the wormhole spacetime for different $\lambda$ and $u_0$. Black curves ($|l| > |l_{\text{ph}}|$) cross the throat, blue curves ($|l| < |l_{\text{ph}}|$) are deflected back, and red curves ($|l| = |l_{\text{ph}}|$) correspond to the circular orbit.}
\end{figure}

To illustrate how the wormhole parameters affect the motion of light rays, we plot the photon trajectories in the polar coordinates $(u, \phi)$ for several representative values of $\lambda$ and $u_0$.  
Figure~\ref{AllTraj} shows a set of such trajectories, obtained by numerically integrating the geodesic equations (\ref{t_GS})--(\ref{phi_GS}) with $\mu = 0$. Here, we distinguish three types of rays, as in the embedding diagrams.

Comparing the different panels of Fig.~\ref{AllTraj} reveals two clear trends. 
First, increasing the throat length $\lambda$ shifts the circular orbit outward and expands the region occupied by the deflected (blue) trajectories.  
Second, increasing $u_0$ shifts the circular orbit back toward the throat and also reduces the number of black trajectories; that is, a smaller fraction of photons with the same impact parameter $l$ are able to cross the throat.

Before we consider the parameter dependence of the circular orbits, it is important to note the overall structure of photon spheres in this spacetime.  
Because the function $N(u)$ is symmetric, a circular photon orbit always exists at the throat $u=0$, with the impact parameter $l_{\text{th}} = r(0)/N(0)$.  
In addition, depending on the values of $\lambda$ and $u_0$, two more circular orbits can appear symmetrically on either side of the throat: one at $u_{\text{ph}} > 0$ and its mirror image at $u_{\text{ph}} < 0$.  
These outer orbits emerge when $\lambda$ is sufficiently large or $u_0$ is sufficiently small.
However, the inner throat orbit does not affect the shadow boundary.  
Since its impact parameter $|l_{\text{th}}|$ is larger than $|l_{\text{ph}}|$ for the outer photon sphere (if the latter exists), any photon arriving from infinity with $|l| = |l_{\text{th}}|$ has a turning point before it can reach even the outer orbit.  

\begin{figure}[h!]
    \begin{minipage}[h]{0.4\linewidth}
        \center{\includegraphics[width=1\linewidth]{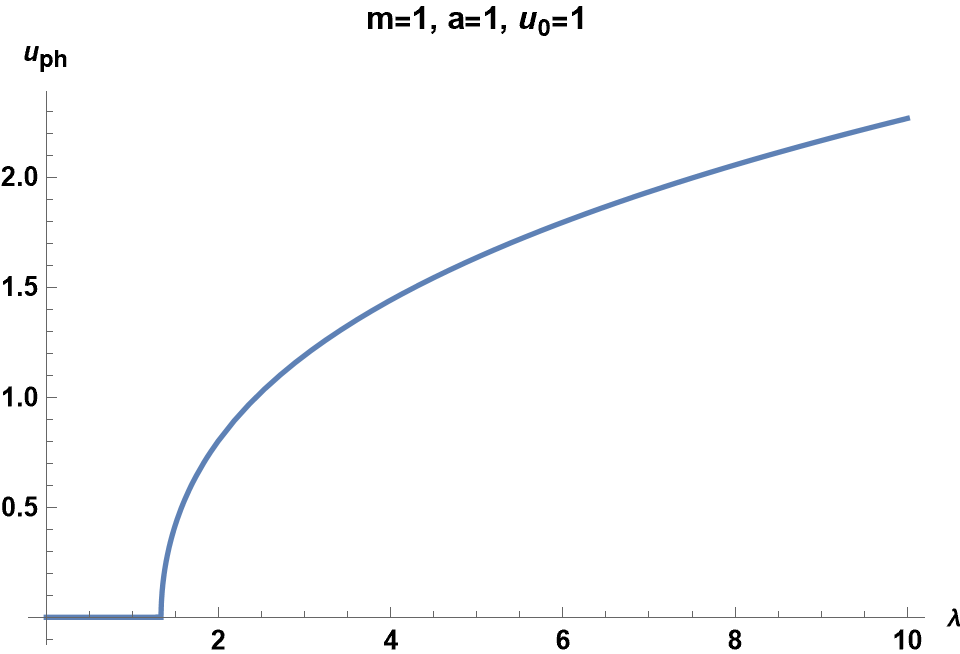}}
    \end{minipage}
    \hspace{1.5cm}
    \begin{minipage}[h]{0.4\linewidth}
        \center{\includegraphics[width=1\linewidth]{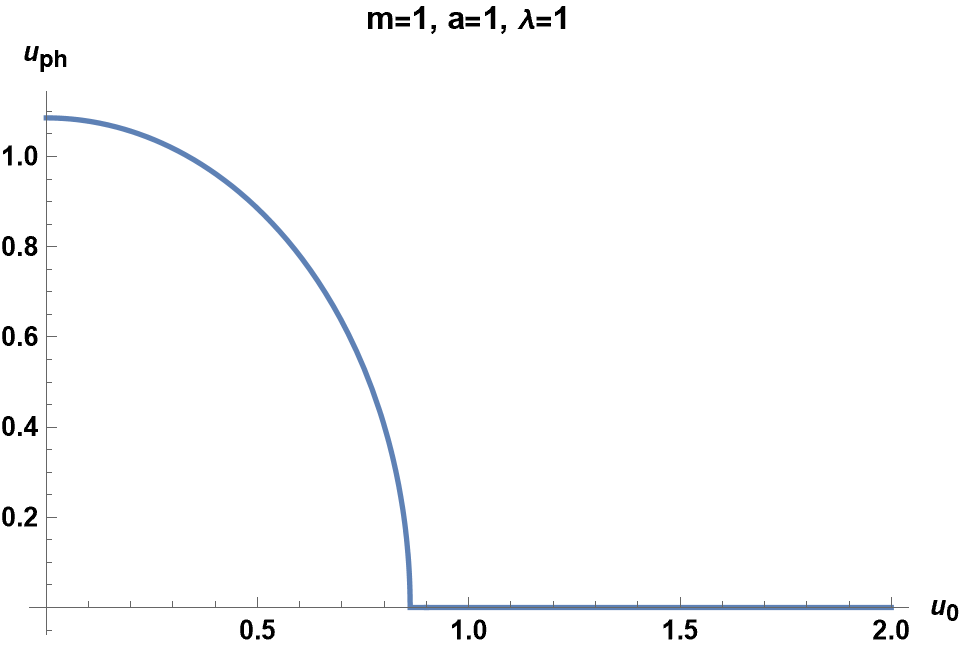}}
    \end{minipage}
    \caption{\label{uph}Location of the circular orbit $u_{\text{ph}}$ as a function of $\lambda$ (left) and $u_0$ (right).}
\end{figure}

The location of the outer circular orbit $u_{\text{ph}}$ (on the observer's side) depends on both $\lambda$ and $u_0$.  
Figure~\ref{uph} shows this dependence, obtained by solving Eqs.~(\ref{Uph_GS})--(\ref{U'ph_GS}) numerically.  
For small $\lambda$, the circular orbit lies at the throat., so that $u_{\text{ph}} = 0$.  
As $\lambda$ increases, $u_{\text{ph}}$ grows and gradually moves outward.  
Increasing the parameter $u_0$, on the other hand, \textit{decreases} the position of the circular orbit $u_{\text{ph}}$, pushing it back toward the throat.

\subsection{Shadow and throat silhouette}

The presence of a wormhole can be found from a dark spot in its observed image.  
This spot can be either the shadow, if there is a luminous background behind the wormhole, or the throat silhouette, if the photons emitted near the throat.
Both quantities were derived in full generality in Secs.~\ref{subsec:shadow} and~\ref{subsec:silhouette}.  
For convenience, we recall the final expressions:
\begin{equation}
    \alpha_{\text{sh}} = |l_{\text{ph}}| = \left|\frac{r(u_{\text{ph}})}{N(u_{\text{ph}})}\right|,
    \qquad
    \int_0^\infty \frac{du}{r^2(u) \sqrt{N^2(u) - \alpha_{\text{sil}}^2/r^2(u)}} = \frac{\pi}{\alpha_{\text{sil}}}.
    \label{shadow_sil_def}
\end{equation}
Here $u_{\text{ph}}$ is the location of the outermost (unstable) photon sphere, determined by Eqs.~(\ref{Uph_GS})--(\ref{U'ph_GS}), and $\alpha_{\text{sil}}$ is the angular radius of the throat silhouette.

\begin{figure}[h!]
    \begin{minipage}[h]{0.48\linewidth}
        \centering
        \includegraphics[width=\linewidth]{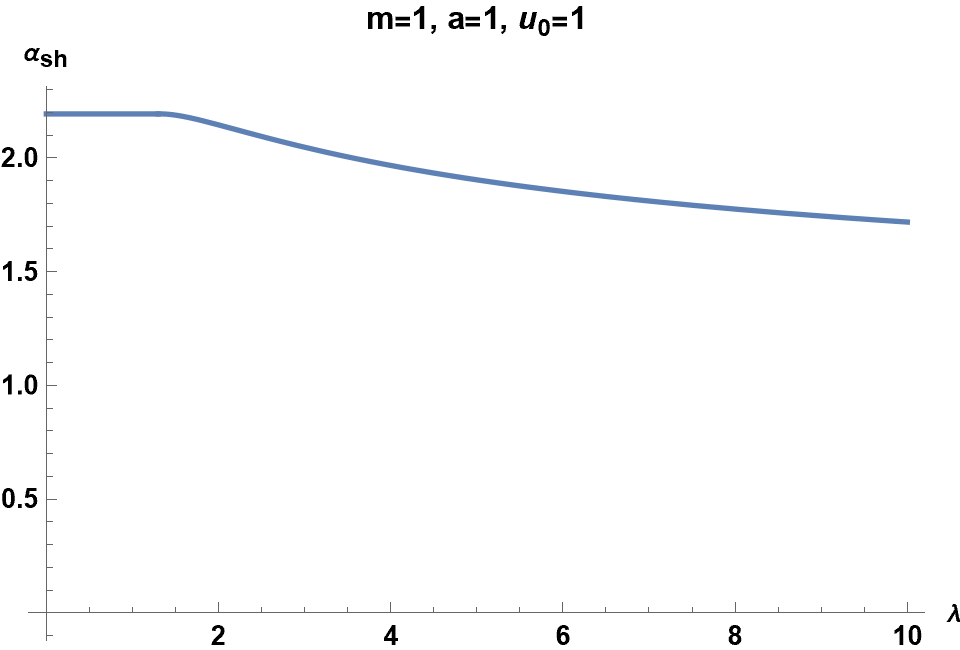}\\
        \includegraphics[width=\linewidth]{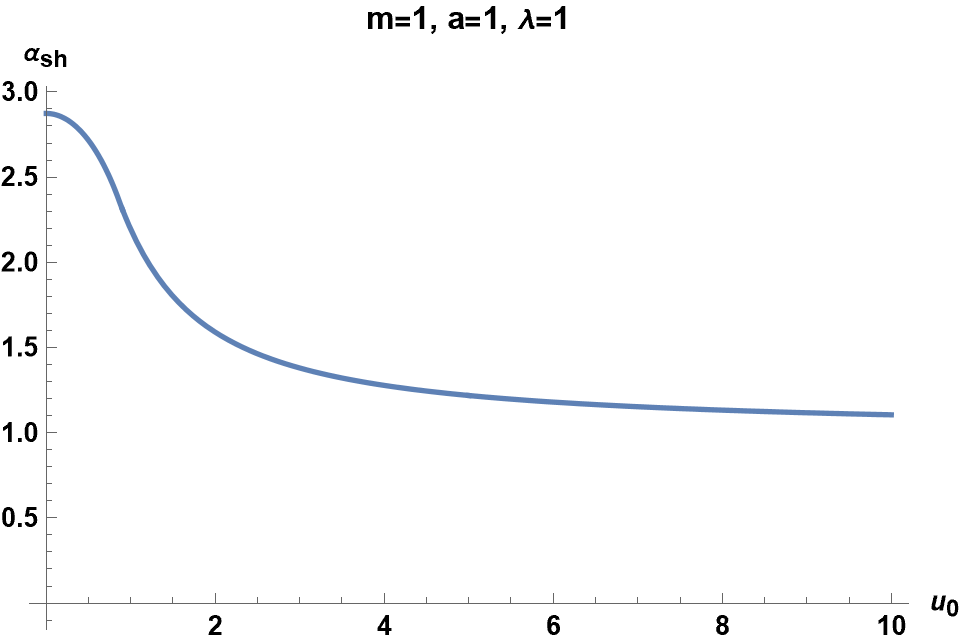}
    \end{minipage}
    \hfill
    \begin{minipage}[h]{0.48\linewidth}
        \centering
        \includegraphics[width=\linewidth]{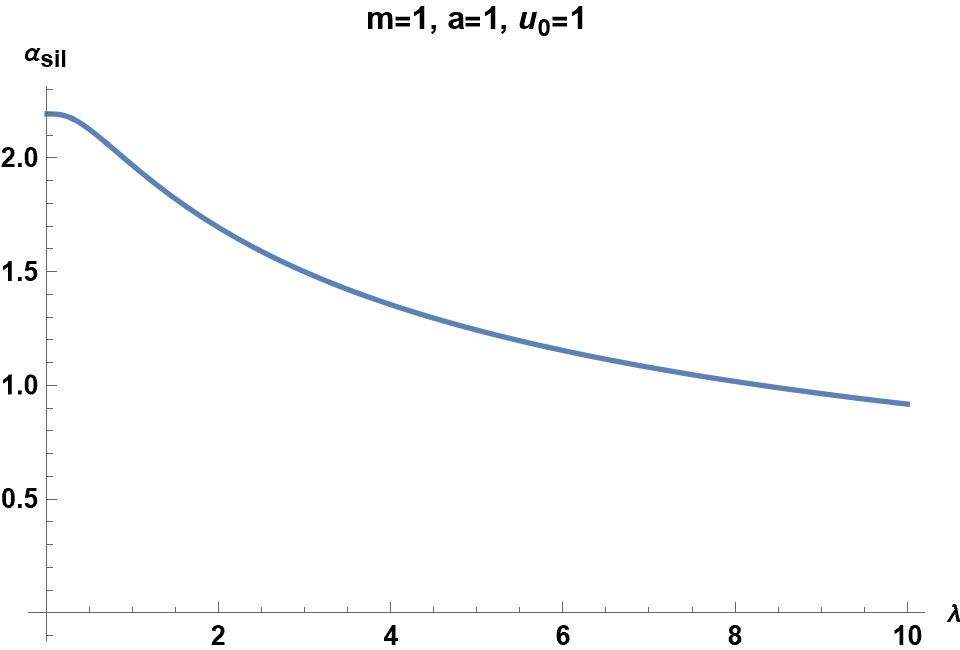}\\
        \includegraphics[width=\linewidth]{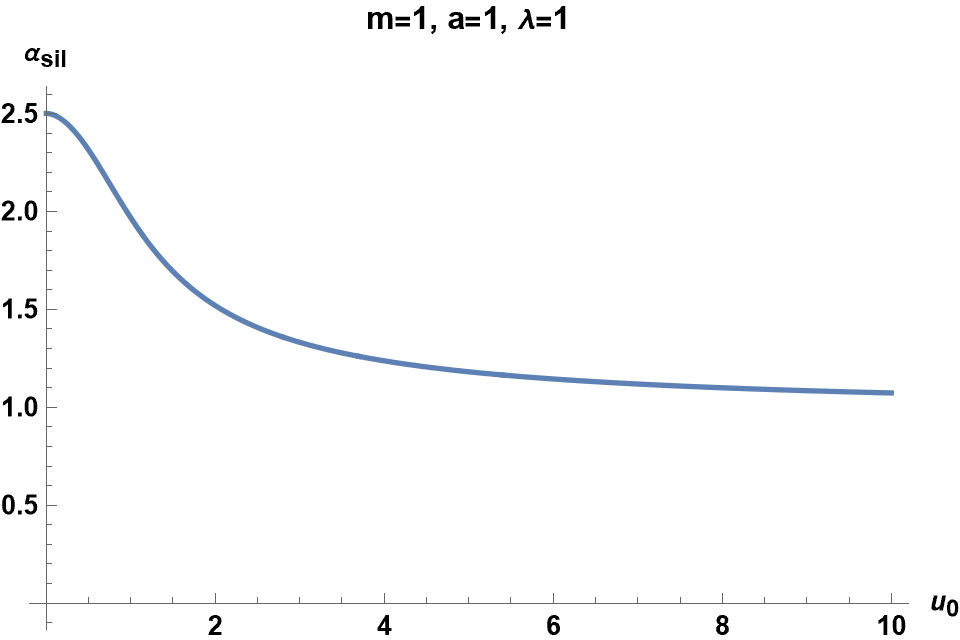}
    \end{minipage}
    \caption{\label{shad_sil_param}Shadow radius $\alpha_{\text{sh}}$ (left column) and throat silhouette radius $\alpha_{\text{sil}}$ (right column) as functions of $\lambda$ (top) and $u_0$ (bottom).}
\end{figure}

We first consider the dependence of both radii on the parameters $\lambda$ and $u_0$. The results are shown in Fig.~\ref{shad_sil_param}.  
The shadow and the silhouette behave very similarly: each radius decreases as either $\lambda$ or $u_0$ becomes larger.   
For sufficiently small $\lambda$ or large $u_0$, the circular orbit is at the throat, and the shadow radius is $\alpha_{\text{sh}} = a / N(0)$.

\begin{figure}[h!]
    \begin{minipage}[h]{0.48\linewidth}
        \centering
        \includegraphics[width=\linewidth]{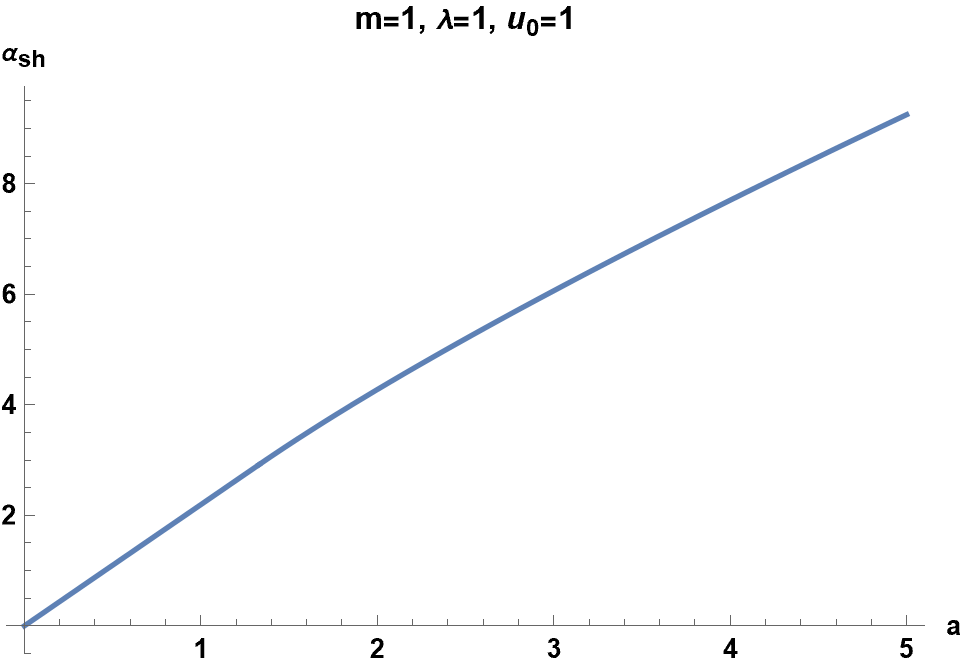}
        \caption{\label{shad_a}Shadow radius $\alpha_{\text{sh}}$ as a function of the throat radius $a$.}
    \end{minipage}
    \hfill
    \begin{minipage}[h]{0.48\linewidth}
        \centering
        \includegraphics[width=\linewidth]{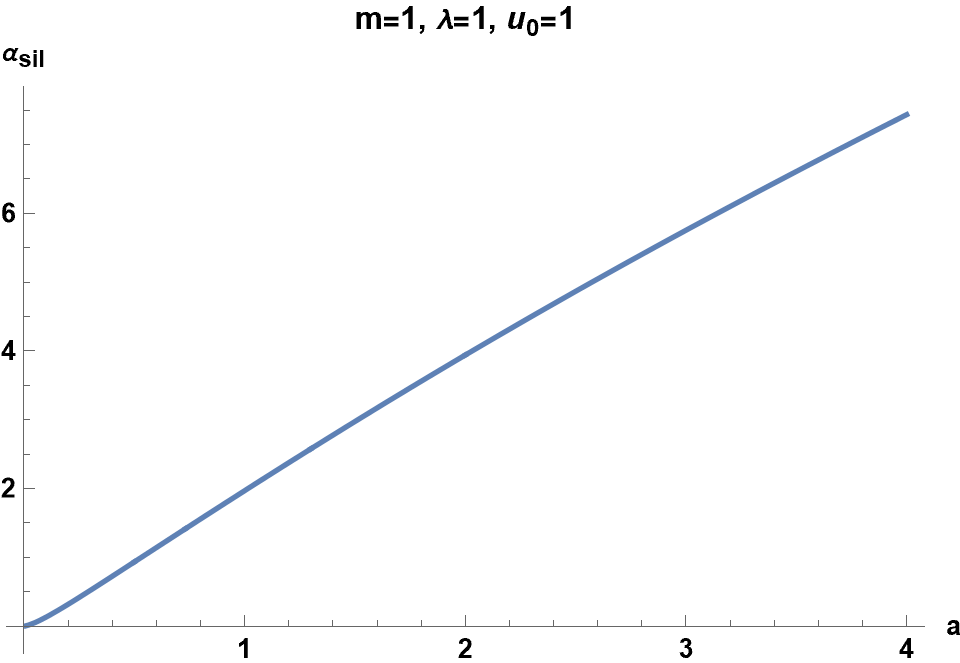}
        \caption{\label{silhouette_a}Throat silhouette radius $\alpha_{\text{sil}}$ as a function of the throat radius $a$.}
    \end{minipage}
\end{figure}

In our wormhole model, besides $\lambda$ and $u_0$, there is one more free parameter -- the throat radius $a$.  
From Figs.~\ref{shad_a} and~\ref{silhouette_a} we see that both $\alpha_{\text{sh}}$ and $\alpha_{\text{sil}}$ increases with increasing $a$.  
For this reason, we can not distinguish the wormhole from a Schwarzschild black hole of equal mass $m$ withe the help of the shadow and throat silhouette radii.
For example, for a Schwarzschild black hole, the shadow and event horizon silhouette radii are $\alpha_{\text{sh}}^{\text{Schw}} = 3\sqrt{3}\,m \approx 5.196\,m$ and $\alpha_{\text{sil}}^{\text{Schw}} \approx 4.457\,m$ \cite{Dokuchaev:2019jqq}.
In our wormhole model, the same shadow radius is obtained for $a \approx 2.5$, $u_0 = 1$, $\lambda = 1$, and the same throat silhouette radius for $a \approx 2.27$, $u_0 = 1$, $\lambda = 1$.  

\newpage
\subsection{Images of an accretion disk}
We now apply the procedure outlined in Sec.~\ref{sec:accretion} to construct the images of an accreting wormhole and compare them with that of a Schwarzschild black hole of the same mass. The observer is located at $u_o = 10\,000$ and at a polar angle $\theta_o = 84.24^\circ$, a value close to the orientation of Sgr A$^*$ with respect to the Earth. All objects have the same mass $m=1$, and the throat radii of the wormhole models are chosen as $a=2$, which coincides with the Schwarzschild event horizon radius $r_{\rm h}=2m$. Despite these coincidences, the resulting images are noticeably different.

For each configuration we solve the integral equations (\ref{noturnpoint})--(\ref{turnpoint}) for photons emitted in the equatorial plane ($\theta_s=\pi/2$) from the region $0\leq u_s\leq u_{\rm ISCO}$ and reaching the observer. The redshift factor $g(l,k)=1/\varepsilon(l,k)$ is obtained from Eq.~(\ref{photon_energy}) and is used to colour every point on the celestial plane $(\alpha,\beta)$. The colour scale ranges from dark red ($g\approx 0$, strongly redshifted photons) to bright yellow ($g\gtrsim 1$, weakly redshifted or blueshifted photons). The corresponding numerical values of the celestial coordinates and the energy shift are collected in Table~\ref{tab:image_data}.
\begin{table}[h!]
\centering
\caption{Ranges of the celestial coordinates $\alpha$, $\beta$ and the energy shift factor $g$ for the Schwarzschild black hole image and for the wormhole images with three different sets of parameters. For each object the values at the horizon/throat ($r=2$ or $u=0$) and at the innermost stable circular orbit (ISCO) are given.}
\label{tab:image_data}
\begin{tabular}{|l| l| c c| c c |c c|}
\hline
Object & Region & $\alpha_{\min}$ & $\alpha_{\max}$ & $\beta_{\min}$ & $\beta_{\max}$ & $g_{\min}$ & $g_{\max}$ \\
\hline
Schwarzschild BH & $r_{\rm h}=2m$    & $-3.85$  & $3.85$  & $-0.032$ & $4.327$ & $0$ & $0$ \\
                  & $r_{\rm isco}=6m$     & $-7.41$  & $7.41$  & $-0.643$ & $5.343$ & $0.582$                & $0.742$ \\
\hline
WH $\lambda{=}1, u_0{=}1$  & $u_{\rm th}{=}0$   & $-3.678$ & $3.678$ & $-0.215$ & $3.906$ & $0.122$ & $0.372$ \\
                          & $u_{\rm isco}\approx 2.97 m$      & $-5.421$ & $5.421$ & $-0.460$ & $4.303$ & $0.369$ & $1.452$ \\
\hline
WH $\lambda{=}1, u_0{=}0.5$ & $u_{\rm th}{=}0$  & $-4.044$ & $4.044$ & $-0.229$ & $4.306$ & $0.061$ & $0.101$ \\
                            & $u_{\rm isco}\approx 3.76 m$     & $-6.156$ & $6.156$ & $-0.539$ & $4.654$ & $0.413$ & $1.436$ \\
\hline
WH $\lambda{=}4, u_0{=}1$   & $u_{\rm th}{=}0$  & $-2.657$ & $2.657$ & $-0.133$ & $2.885$ & $0.121$ & $0.178$ \\
                            & $u_{\rm isco}\approx 5.37 m$     & $-4.972$ & $4.972$ & $-0.418$ & $3.696$ & $0.529$ & $1.312$ \\
\hline
\end{tabular}
\end{table}

\begin{figure}[h!]
	\begin{minipage}[h]{0.49\linewidth}
		\center{\includegraphics[width=1\linewidth]{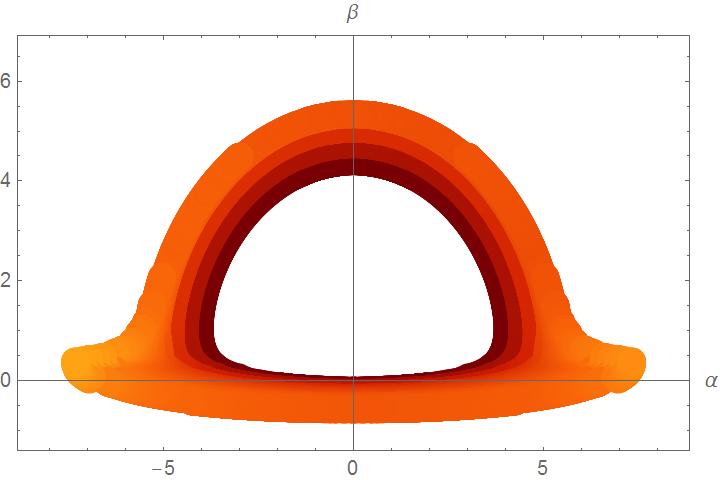}}
	\end{minipage}
	\hfill
	\begin{minipage}[h]{0.49\linewidth}
		\center{\includegraphics[width=1\linewidth]{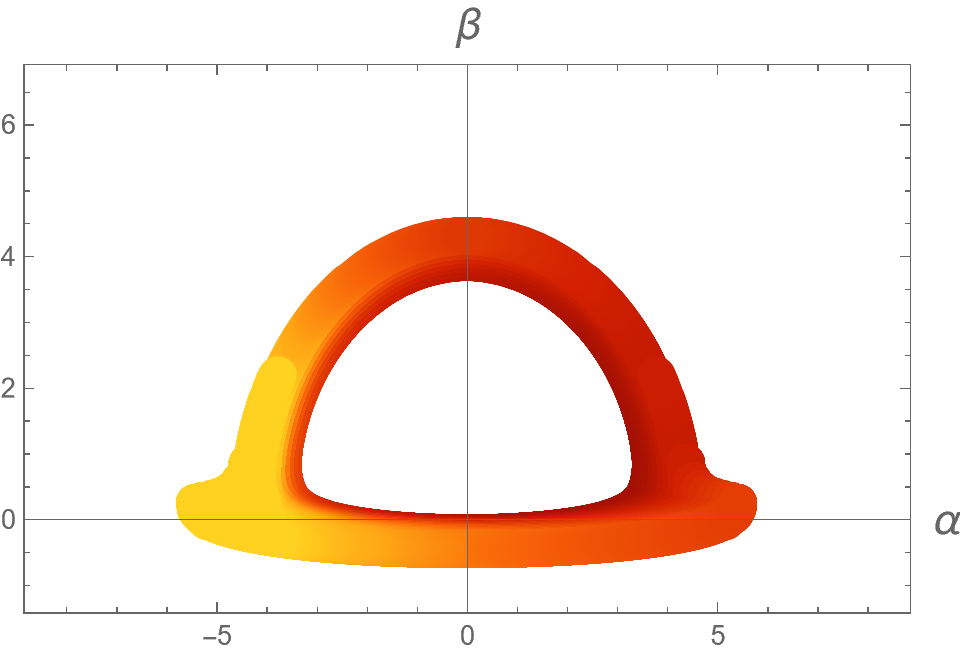}}
	\end{minipage}
	\vfill
	\begin{minipage}[h]{0.49\linewidth}
		\center{\includegraphics[width=1\linewidth]{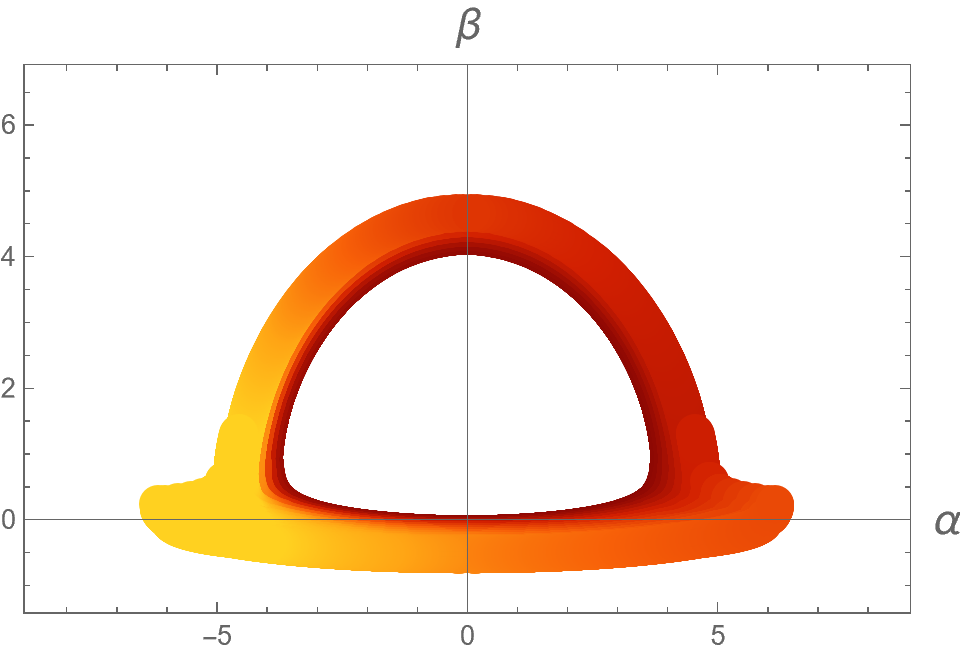}}
	\end{minipage}
	\hfill
	\begin{minipage}[h]{0.49\linewidth}
		\center{\includegraphics[width=1\linewidth]{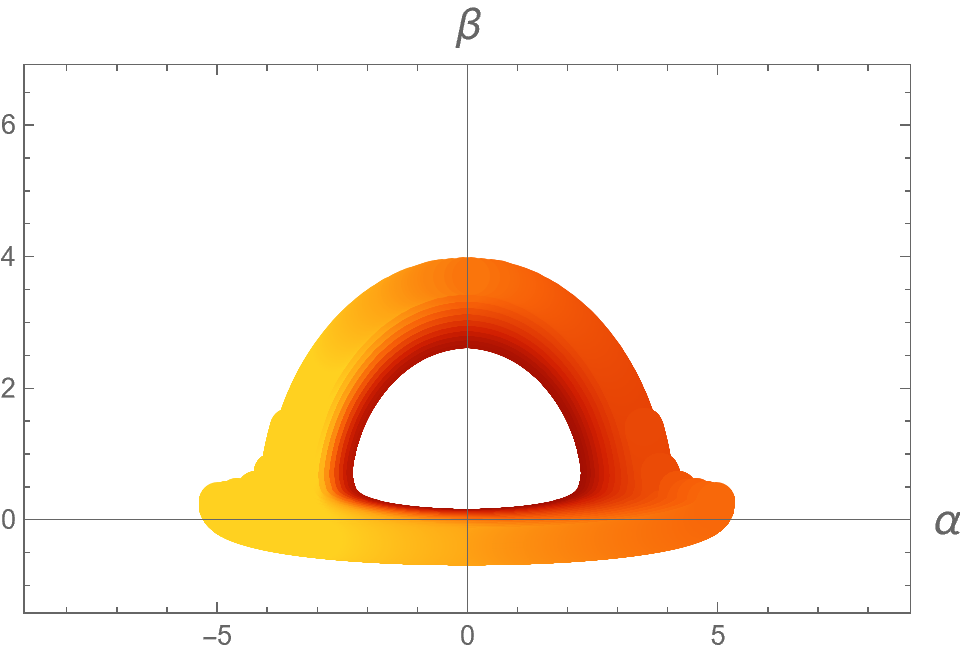}}
	\end{minipage}
	\caption{\label{images}Images of the accreting Schwarzschild black hole (top left) and of the wormhole with $a=2$, $m=1$ and three sets of $(\lambda, u_0)$: $\lambda=1, u_0=1$ (top right), $\lambda=1, u_0=0.5$ (bottom left), $\lambda=4, u_0=1$ (bottom right). The colour indicate the energy shift factor $g$ from dark red ($g\approx0$) to bright yellow ($g\gtrsim1$). The central dark spot is the event horizon silhouette (Schwarzschild) or the throat silhouette (wormhole). The observer is located at $\theta_o=84.24^\circ$.}
\end{figure}

Figure~\ref{images} shows the images for the Schwarzschild black hole and for three wormhole configurations with different sets of $(\lambda, u_0)$: $(\lambda=1, u_0=1)$, $(\lambda=1, u_0=0.5)$, and $(\lambda=4, u_0=1)$. The central \textit{dark} region is the silhouette of the event horizon (for the black hole) or the silhouette of the throat (for the wormholes), formed by photons that emitted near the compact object.

The size of the dark silhouette is most easily quantified by its vertical coordinate $\beta$. For the Schwarzschild black hole the event horizon silhouette is bounded by $\beta_{\rm max}\approx 4.33$ and $\beta_{\rm min}\approx -0.03$. For the wormhole with $(\lambda=1, u_0=1)$ we have $\beta_{\rm max}\approx 3.91$, which is noticeably smaller than the Schwarzschild value. Making the gravitational well deeper by decreasing $u_0$ to $0.5$ increases the silhouette; in this case $\beta_{\rm max}\approx 4.31$ is almost equal to that of the black hole. And, vice versa, elongating the throat by increasing $\lambda$ to $4$ decreases the silhouette to $\beta_{\rm max}\approx 2.89$. These trends are completely agreement with the analysis of previously section, where we showed that the throat silhouette radius grows with decreasing $u_0$ and decreasing $\lambda$. It should be noted that the horizontal coordinate $\alpha$ of the silhouette behaves similarly.

As for the energy shift, photons emitted near the event horizon of the Schwarzschild black hole experience very strong redshift, with $g\approx 0$, making the event horizon silhouette black.
Photons emitted from the wormhole throat have noticeably larger $g$: $g_{\rm min}\approx 0.122$ for $(\lambda=1, u_0=1)$, $g_{\rm min}\approx 0.061$ for $(\lambda=1, u_0=0.5)$, and $g_{\rm min}\approx 0.121$ for $(\lambda=4, u_0=1)$. At the innermost stable circular orbit the maximum value of $g$ is significantly larger for the wormholes ($g_{\rm max}\approx 1.45$ for $\lambda=1, u_0=1$, $g_{\rm max}\approx 1.44$ for $\lambda=1, u_0=0.5$) and ($g_{\rm max}\approx 1.31$ for $\lambda=4, u_0=1$ than for the Schwarzschild black hole ($g_{\rm max}\approx 0.74$), because the Doppler effect plays a major role. As a result, the innermost part of the wormhole accretion disk appears significantly brighter than that for the black hole.

Let's also consider the effect of wormhole parameters $\lambda$ and $u_0)$ on the photon energy shift.  A deeper gravitational well (smaller $u_0$) strongly decreases $g$ at the throat, when the values at the ISCO remain nearly unchanged. A longer throat (larger $\lambda$) leaves $g_{\min}$ at the throat almost unaffected but noticeably decreases $g_{\max}$; at the ISCO it raises $g_{\min}$ and lowers $g_{\max}$, making the energy shift distribution more homogeneous. Thus, the parameters $\lambda$ and $u_0$ provide independent control over both the size of the dark spot and the brightness distribution in the accretion disk image.

In summary, the combined analysis of the silhouette size and the energy shift distribution provides a reliable instrument to distinguish a wormhole from a black hole. 

\begin{acknowledgments}
This work was supported by the Theoretical Physics and Mathematics Advancement Foundation ``BASIS'', grant No.~24-1-1-39-2.
\end{acknowledgments}


\begin{thebibliography}{99}
	
	\bibitem{Flamm} 
	L.~Flamm, 
	``Beitr\"age zur Einsteinschen Gravitationstheorie,'' 
	Phys. Z. \textbf{17}, 448 (1916).
	
	\bibitem{EinsteinRosen}
	A.~Einstein and N.~Rosen, 
	``The Particle Problem in the General Theory of Relativity,'' 
	Phys. Rev. \textbf{48}, 73-77 (1935).
	
	\bibitem{Wheeler_Geons} 
	J.~A.~Wheeler, 
	``Geons,'' 
	Phys. Rev. \textbf{97}, 511–536 (1955).
	
	\bibitem{Wheeler_GDs} 
	J.~A.~Wheeler, 
	\textit{Geometrodynamics} (Academic Press, New York, 1962, 334p.).
	
	\bibitem{MorrisThorne1}
	M.~S.~Morris and K.~S.~Thorne, 
	``Wormholes in spacetime and their use for interstellar travel: A tool for teaching general relativity,'' 
	Am. J. Phys. \textbf{56}, 395-412 (1988).
	
	\bibitem{VisserBook} 
	M.~Visser, 
	\textit{Lorentzian Wormholes: From Einstein to Hawking} (American Institute of Physics, Woodbury, 1995, 412p.).
	
	\bibitem{LoboReview} 
	F.~S.~N.~Lobo, 
	``Exotic solutions in General Relativity: Traversable wormholes and ``warp drive'' spacetimes,'' 
	in \textit{Classical and Quantum Gravity Research} (Nova Science Publishers, New York, 2008), pp.~1--78.
	
	\bibitem{BambiStojkovic}
	C.~Bambi and D.~Stojkovic, 
	``Astrophysical Wormholes,'' 
	Universe \textbf{7}, 136 (2021).
	
	\bibitem{Bronnikov:2023lza}
	K.~A.~Bronnikov and S.~V.~Sushkov,
	``Current problems and recent advances in wormhole physics'' 
	Universe \textbf{9}, 39 (2023).
	
	\bibitem{EHT} 
	The Event Horizon Telescope Collaboration, 
	``First M87 Event Horizon Telescope Results. I. The Shadow of the Supermassive Black Hole,'' 
	Astrophys. J. Lett. \textbf{875}, L1 (2019).
	
	\bibitem{TsuHarYaj}
	N.~Tsukamoto, T.~Harada, and K.~Yajima, 
	``Can we distinguish between black holes and wormholes by their Einstein-ring systems?'' 
	Phys. Rev. D \textbf{86}, 104062 (2012).
	
	\bibitem{Perlick}
	V.~Perlick, 
	``Exact gravitational lens equation in spherically symmetric and static spacetimes,'' 
	Phys. Rev. D \textbf{69}, 064017 (2004).
	
	\bibitem{NanZhaZak}
	K.~K.~Nandi, Y.-Z.~Zhang, and A.~V.~Zakharov, 
	``Gravitational lensing by wormholes,'' 
	Phys. Rev. D \textbf{74}, 024020 (2006).
	
	\bibitem{Muller}
	T.~M\"uller, 
	``Exact geometric optics in a Morris-Thorne wormhole spacetime,'' 
	Phys. Rev. D \textbf{77}, 044043 (2008).
	
	\bibitem{Shaikh:2019jfr}
	R.~Shaikh, P.~Banerjee, S.~Paul, and T.~Sarkar, 
	``Strong gravitational lensing by wormholes,'' 
	JCAP \textbf{07}, 028 (2019) 
	[erratum: JCAP \textbf{12}, E01 (2023)].
	
	\bibitem{Bronnikov:2018nub}
	K.~A.~Bronnikov and K.~A.~Baleevskikh, 
	``On gravitational lensing by symmetric and asymmetric wormholes,'' 
	Grav. Cosmol. \textbf{25}, 44 (2019).
	
	\bibitem{Shaikh:2018oul}
	R.~Shaikh, P.~Banerjee, S.~Paul, and T.~Sarkar, 
	``A novel gravitational lensing feature by wormholes,'' 
	Phys. Lett. B \textbf{789}, 270 (2019) 
	[erratum: Phys. Lett. B \textbf{791}, 422 (2019)].
	
	\bibitem{Ishkaeva:2023xny}
	V.~A.~Ishkaeva and S.~V.~Sushkov, 
	``Image of an accreting general Ellis-Bronnikov wormhole,'' 
	Phys. Rev. D \textbf{108}, 084054 (2023).
	
	\bibitem{Pal:2023}
	S.~Pal, K.~Pal, R.~Shaikh, and T.~Sarkar,
	``A rotating modified JNW spacetime as a Kerr black hole mimicker,''
	J. Cosmol. Astropart. Phys. \textbf{11}, 060 (2023)
	
	\bibitem{Wielgus_etal:2020}
	M.~Wielgus, J.~Hor\'ak, F.~Vincent, and M.~Abramowicz, 
	``Reflection-asymmetric wormholes and their double shadows,'' 
	Phys. Rev. D \textbf{102}, 084044 (2020).
	
	\bibitem{Shaikh:2018kfv}
	R.~Shaikh, 
	``Shadows of rotating wormholes,'' 
	Phys. Rev. D \textbf{98}, 024044 (2018).
	
	\bibitem{Amir:2018pcu}
	M.~Amir, K.~Jusufi, A.~Banerjee, and S.~Hansraj, 
	``Shadow images of Kerr-like wormholes,'' 
	Class. Quant. Grav. \textbf{36}, 215007 (2019).
	
		\bibitem{Guerrero:2021}
	M.~Guerrero, G.~J.~Olmo, and D.~Rubiera-Garcia, 
	``Double shadows of reflection-asymmetric wormholes supported by positive energy thin-shells,'' 
	JCAP \textbf{04}, 066 (2021).
	
	
	\bibitem{Gjorgjieski:2025uik}
	K.~Gjorgjieski, J.~Kunz, and P.~Nedkova, 
	``Circular orbits and photon orbits at wormhole throats,'' 
	Phys. Rev. D \textbf{112}, L101501 (2025).
	
	\bibitem{Gjorgjieski:2025matter}
	K.~Gjorgjieski, J.~Kunz, and P.~Nedkova, 
	``Matter accumulations and accretion tori around wormholes,'' 
	Phys. Rev. D \textbf{112}, 064008 (2025).
	
	\bibitem{Tan:2025cte}
	K.~Tan and X.~G.~Lan,
	``Asymmetric thin-shell wormholes in the Kalb-Ramond background: Observational characteristics and extra photon rings,''
	Chin. Phys. C \textbf{50} (2026) no.4, 045101
	
	\bibitem{Novikov:2025elo}
	I.~D.~Novikov, S.~V.~Repin and D.~A.~Paksivatova,
	``Observing an accretion disk inside a wormhole shadow,''
	[arXiv:2508.02752 [gr-qc]].
	
	\bibitem{Macedo:2025ipc}
	C.~F.~B.~Macedo, J.~L.~Rosa, D.~Rubiera-Garcia and A.~Rueda,
	``Multiphoton ring structure of reflection-asymmetric traversable thin-shell wormholes,''
	Phys. Rev. D \textbf{113} (2026) no.6, 064004
	
	\bibitem{Bambhaniya:2021ugr}
	P.~Bambhaniya, S.~K, K.~Jusufi and P.~S.~Joshi,
	``Thin accretion disk in the Simpson-Visser black-bounce and wormhole spacetimes,''
	Phys. Rev. D \textbf{105} (2022) no.2, 023021
	
	
	
	\bibitem{BardeenPress}
	J.~M.~Bardeen, W.~H.~Press, and S.~A.~Teukolsky, 
	``Rotating black holes: Locally nonrotating frames, energy extraction, and scalar synchrotron radiation,'' 
	Astrophys. J. \textbf{178}, 347 (1972).
	
	\bibitem{PerlickTsupko}
	V.~Perlick and O.~Y.~Tsupko,
	``Calculating black hole shadows: Review of analytical studies,''
	Phys. Rept. \textbf{947}, 1-39 (2022)
	
	\bibitem{Dokuchaev:2019jqq}
	V.~I.~Dokuchaev and N.~O.~Nazarova, 
	``Silhouettes of invisible black holes,'' 
	Usp. Fiz. Nauk \textbf{190}, 627 (2020) 
	[Phys. Usp. \textbf{63}, 583 (2020)].
	
	\bibitem{SushkovDwalls}
	S.~V.~Sushkov, 
	``Domain walls in wormhole space-time,'' 
	Grav. Cosmol. \textbf{7}, 197 (2001).
	
	\bibitem{Popov:2018}
	A.~A.~Popov,
	``Semiclassical long throats of the wormholes,''
	arXiv:1809.06202 [hep-th] (2018).
	
	
	\bibitem{Carter}
	B.~Carter, 
	``Global Structure of the Kerr Family of Gravitational Fields,'' 
	Phys. Rev. \textbf{174}, 1559 (1968).
	
	\bibitem{Vasquez}
	S.~V\'{a}squez and E.~Esteban, 
	``Strong field gravitational lensing by a Kerr black hole,'' 
	Nuovo Cim. B \textbf{119}, 489 (2004).
	
\end{thebibliography}
\end{document}